\documentclass[conference]{IEEEtran}

\usepackage[T1]{fontenc}
\usepackage[utf8]{inputenc}

\usepackage{booktabs} 
\usepackage{tabulary} 
\usepackage{collcell}
\usepackage{enumitem}

\usepackage{cancel}
\usepackage[final]{listings} 
\usepackage{msc}
\usepackage{standalone,tikz}
\usepackage[singlelinecheck=true,justification=raggedright]{subcaption}
\usepackage[fixed]{fontawesome5}

\usepackage{xifthen}
\usepackage{xspace}   

\usepackage{cite}
\usepackage{amsmath,amssymb,amsfonts}
\usepackage{algorithmic}
\usepackage{graphicx}
\usepackage{textcomp}
\usepackage{xcolor}
\def\BibTeX{{\rm B\kern-.05em{\sc i\kern-.025em b}\kern-.08em
	T\kern-.1667em\lower.7ex\hbox{E}\kern-.125emX}}

\usepackage{hyperref}

\newcommand*{\doi}[1]{\nolinkurl{doi:#1}}
\makeatletter
\g@addto@macro{\UrlBreaks}{\do\*\do\-\do\~\do\'\do\"\do\-}
\makeatother

\newcommand*{\sref}[1]{\hyperref[#1]{\S~\ref*{#1}}}
\newcommand*{\aref}[1]{\hyperref[#1]{Appendix~\ref*{#1}}}

\definecolor{i-cyan}{RGB}{0,157,224}
\definecolor{i-dark}{RGB}{70,85,95}
\colorlet{code}{i-cyan!25!i-dark}
\definecolor{pass}{HTML}{3C8031} 
\definecolor{fail}{HTML}{A9341F} 

\newcommand*{\var}[1]{\mathit{#1}}

\newcommand*{\hera}{Hera\xspace}
\newcommand*{\hema}{Hema\xspace}

\newcommand*{\tamarinresult}[1]{
	\ifthenelse{\isin{v}{#1} \OR \isin{f}{#1}}{
		\ifthenelse{\isin{v}{#1}}{
			\textcolor{pass}{#1}
		}{
			\textcolor{fail}{#1}
		}
	}{
		#1
	}
}
\newcolumntype{T}{>{\collectcell\tamarinresult}c<{\endcollectcell}}

\usetikzlibrary{arrows.meta}
\usetikzlibrary{backgrounds}
\usetikzlibrary{calc}
\usetikzlibrary{decorations.text}
\usetikzlibrary{fit}
\usetikzlibrary{patterns}
\usetikzlibrary{positioning}
\usetikzlibrary{shadows}
\usetikzlibrary{shapes.misc}

\tikzset{
	estado estilo/.style={node distance=1.2cm},
	pics/estado/.style n args={3}{
		code = {
			\begin{scope}[
				node distance=0cm, outer sep=0cm,
				cell/.style={draw, minimum height=0.6cm},
				every node/.append style={anchor=center}
			]
				\path node[cell] (-a) [minimum width=1cm, align=center] {#1} node[cell] (-b) [minimum width=4cm, align=center, right=of -a] {#2} let \p1=($(-b.south east)-(-a.south west)$), \n1={veclen(\p1)} in node[cell] (-c) [below=of -a.south west, anchor=north west, minimum width=\n1, text width=(\pgfkeysvalueof{/pgf/minimum width}-2*\pgfkeysvalueof{/pgf/inner xsep}), align=left] {#3} node (-g) [fit=(-a)(-b)(-c), inner sep=0pt] {};
			\end{scope}
		}
	},
	estado seta/.style={-{stealth}}
}

\newcommand*\numd[1]{\tikz[baseline=(char.base)]{\node [solid, rounded rectangle, minimum size=0.3cm, inner sep=0cm, text height=1ex, text depth=0.1ex, fill=magenta!25!black, text=white, font=\scriptsize] (char) {#1}}}

\mscset{
	action width=10em,
	action height=2.7ex,
	every action/.style={inner sep=0.1cm},
}

\lstdefinestyle{text}{
	basicstyle       = \normalfont,
	columns          = fullflexible,
	breaklines       = true,
	literate         = {<-}{{${}\leftarrow{}$}}{2}{->}{{${}\rightarrow{}$}}{2},
}

\lstdefinestyle{code}{
	basicstyle       = \normalfont\color{code},
	columns          = fullflexible,
	breaklines       = true,
	literate         = {<-}{{${}\leftarrow{}$}}{2}{->}{{${}\rightarrow{}$}}{2},
}

\lstdefinestyle{tese}{ 
	breaklines = true,
	frame = l,
	rulecolor = \color{gray!80},
	numbers = left,
	numberblanklines = false,
	numberstyle = \footnotesize,
	firstnumber = last,
	xleftmargin = 2em,
	captionpos = t,
}
\lstdefinestyle{protocolo}{
	style = tese, 
	basicstyle = \ttfamily,
	columns = fixed,
	mathescape = true,
	escapechar = {\%},
	literate = {
		{<-}{${}\leftarrow{}$}{2}
		{->}{$\rightarrow$}{2}
		{A->B}{$A \rightarrow B$}{4} 
	},
}
\lstdefinelanguage{tamarin-output}{
	basicstyle = \normalfont\small,
	columns = fullflexible,
	classoffset = 0,
	morekeywords = {},
	keywordstyle = \bfseries,
	classoffset = 1,
	morekeywords = {falsified},
	keywordstyle = \color{fail}\bfseries,
	classoffset = 2,
	morekeywords = {verified},
	keywordstyle = \color{pass}\bfseries,
	classoffset = 0, 
}

\lstdefinelanguage{tese}{
	morekeywords={
		fresh,
		ra,
		attest,
		verify,
		getpol,
		validate,
		ma,
		hash,
	},
	literate={<-}{{${}\leftarrow{}$}}{2}{->}{{${}\rightarrow{}$}}{2},
}
\lstdefinelanguage{hema}{}
\newcommand*{\fun}{\lstinline[mathescape,language=tese]}
\newcommand*{\lfun}[3]{%
	\ifthenelse{\isempty{#3}}{%
		\textbf{#1}(#2)%
	}{%
		#3 \leftarrow \textbf{#1}(#2)%
	}
}
\newcommand*{\rfun}[3]{%
	\ifthenelse{\isempty{#3}}{%
		\textbf{#1}(#2)%
	}{%
		\textbf{#1}(#2) \rightarrow #3%
	}
}

\begin{document}

\title{\emph{Know Thy Neighbor:} Cross-TEE Mutual Attestation}

\author{%
Daniel Andrade, João N. Silva, Miguel P. Correia\\
INESC-ID, Instituto Superior Técnico, Universidade de Lisboa\\
daniel.andrade@tecnico.ulisboa.pt, joao.n.silva@inesc-id.pt, miguel.p.correia@tecnico.ulisboa.pt
}

\maketitle

\begin{abstract}
Cloud services are composed of multiple heterogeneous distributed components and instances that communicate with one another. This occurs both in applications and services running in traditional execution environments and in trusted applications (TAs) running in trusted execution environments (TEEs).
TA instances use attestation before exchanging information to ensure all parties meet the expected security conditions. The straightforward solution to mutually attesting two TA instances that are willing to communicate is employing remote attestation mechanisms in both directions.
This is typically the case when the two TA instances are running on TEEs of the same type.
In order to support cross-TEE attestation, such an approach, that is, using remote attestation in both directions, would require each TEE type (e.g., SGX, TrustZone) to support the attestation software stack of all other TEE types with which it needs to interact.
A dedicated cross-TEE mutual attestation solution has multiple benefits in terms of efficiency and security.

This paper presents the Heterogeneous Mutual Attestation (Hema) protocol, a formally-verified protocol for the mutual attestation of TA instances running on the same TEE type or on different TEE types.
\end{abstract}

\begin{IEEEkeywords}
Mutual attestation, protocols, security, trusted execution environments.
\end{IEEEkeywords}

\section{Introduction}
\label{sec:introduction}

Cloud computing~\cite{Armbrust:M:CACM:10} benefits companies by reducing upfront infrastructure costs and their maintenance, which are now handled by a third party, and by increasing the availability, elasticity, and scalability of computing resources in comparison to traditional on-premise computing. There is, however, a potential cost in terms of security and privacy because cloud providers have access to the data located in their cloud services. A malicious employee could forge, modify, delete, or leak customer data irrespective of what any privacy policy might say~\cite{Rocha:DCDV:11:6}. In addition, cloud providers may share data with third parties to satisfy law enforcement requests and warrants. These issues are always a risk to the customers of cloud providers even when a cloud provider follows best practices for securing customer computations and data.

Trusted execution environment (TEE) technologies~\cite{Maene:IEEE:TC:18, Singh:M:Queue:21} such as Intel SGX~\cite{McKeen:HASP:13} and Arm TrustZone~\cite{Arm:09:TrustZone} present themselves as a solution to these security risks by providing isolated containers to run computations, and mechanisms to authenticate what is being computed and how, and to securely store application data outside said containers. Attestation gives the owners of the data confidence on what is being computed.

\subsubsection*{Attestation}

Cloud services are composed of multiple distributed components and instances that communicate with one another. This communication occurs both in applications and services running in traditional execution environments and in trusted applications (TAs) running in TEEs.
TAs use attestation to prove, to other parties, that they meet the expected security conditions, namely that the TA is running on a real and up-to-date TEE, has the correct code and initial data, and, similarly to the TEE, is itself up to date.

\begin{figure}
	\centering
	\begin{subfigure}[t]{0.48\linewidth}
		\centering
		\includegraphics[width=\linewidth]{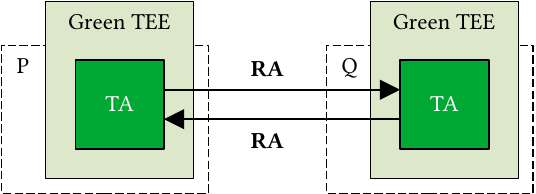}
		\caption{Same TEE type}
		\label{fig:hema:arq:ra-gg}
	\end{subfigure}
	\hfill
	\begin{subfigure}[t]{0.48\linewidth}
		\centering
		\includegraphics[width=\linewidth]{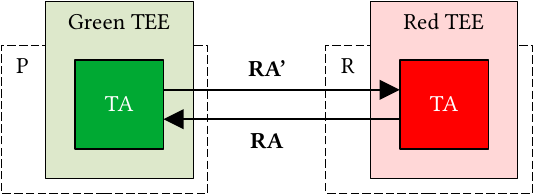}
		\caption{Different TEE types}
		\label{fig:hema:arq:ra-gr}
	\end{subfigure}
	\caption{Mutual attestation by employing remote attestation in both directions. The remote attestation scheme in (\subref{fig:hema:arq:ra-gg}) is the same in both directions, but in (\subref{fig:hema:arq:ra-gr}) the remote attestation schemes are different for each direction (i.e., RA vs RA') because the TEEs hosting the TAs are of different types (which we represented by using different colors, i.e., Green vs Red).}
	\label{fig:hema:arq:ra}
\end{figure}

\begin{figure}
	\centering
	\begin{subfigure}[t]{0.48\linewidth}
		\centering
		\includegraphics[width=\linewidth]{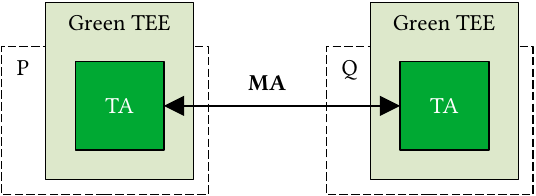}
		\caption{Same TEE type}
		\label{fig:hema:arq:ma-gg}
	\end{subfigure}
	\hfill
	\begin{subfigure}[t]{0.48\linewidth}
		\centering
		\includegraphics[width=\linewidth]{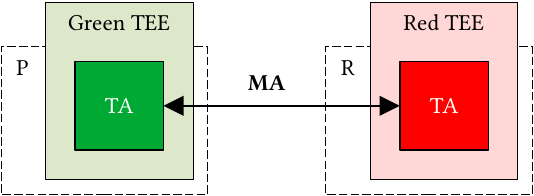}
		\caption{Different TEE types}
		\label{fig:hema:arq:ma-gr-qmark}
	\end{subfigure}
	\caption{Mutual attestation by employing a dedicated cross-TEE mutual attestation solution such as \hera. The mutual attestation scheme remains the same whether the TEEs hosting the TAs have the same type (\subref{fig:hema:arq:ma-gg}) or different types (\subref{fig:hema:arq:ma-gr-qmark}).}
	\label{fig:hema:arq:ma}
\end{figure}

The straightforward solution for attesting two TA instances wanting to communicate is employing remote attestation (RA) mechanisms in both directions as shown in \autoref{fig:hema:arq:ra}. However, this approach could be improved first by using a single software stack for RA and second by having a dedicated mutual attestation (MA) solution as shown in \autoref{fig:hema:arq:ma}.
This is especially true when we combine different TEE types, such as Intel SGX and Arm TrustZone.
Having a single software stack for cross-TEE MA provides the following benefits:

\begin{description}[leftmargin=!,labelwidth=8em,align=right]

	\item[Efficiency] MA is more efficient than using RA twice.
	\item[Security] A single software stack for attestation leads to a smaller Trusted Computing Base (TCB)~\cite{DoD:85:TCSEC}. In addition, using MA instead of RA in both directions is less error prone.

	\item[Correctness] A single software stack for attestation reduces implementation errors in the attestation libraries and on the clients invoking those library.

	\item[Interoperability] Each TEE type supports one MA scheme versus multiple MA schemes. This simplifies compatibility with existing and future TEE types since only one attestation stack is needed.

	\item[Reusability] A single MA stack promotes code reusability.

	\item[Adoption] A single interface for MA benefits both new and old TEE types. New TEE types can adopt the new interface from the beginning. Old TEE types can write a shim to maintain compatibility between the new and the existing interfaces.

\end{description}

A MA scheme is more efficient than using a RA scheme twice since exchanged data can be aggregated in less messages. In addition, for one TEE type to remotely attest another TEE type it needs to support the attestation stack of the Attester. If this same TEE needs to remotely attest N different TEE types then it needs N attestation software stacks. Having multiple attestation software stacks instead of a single one increases (i) the potential for implementation errors leading to more bugs and attack vectors, (ii) the effort required by programmers to maintain compatibility with current TEE types and future TEE types (lower cross-TEE compatibility), and (iii) the size of the TCB (lower security). Finally, a single MA interface and respective software stack promotes code reusability and progressive adoption by existing TEEs.

\subsubsection*{Our Work}

We propose a minimal architecture and protocol, called \emph{Heterogeneous Mutual Attestation} (\hema), for cross-TEE MA enabling two TA instances running on different TEE types to mutually attest themselves. \hema promotes the interoperability and progressive adoption of intra-TEE and inter-TEE MA, and is formally verified using the Tamarin prover~\cite{Tamarin:manual:2405}.
\hema uses a bare-bones heterogeneous RA protocol (\sref{sec:ra}) internally.

\subsubsection*{Applications}

Cross-TEE MA is advantageous in modular and distributed environments.

In modular programming an application is separated into independent modules that interact with one another via well-defined interfaces. This separation makes it easier for large teams to collaborate in the development of a complex application, and promotes the reusability and upgradeability of individual modules. These modules have to authenticate each other before exchanging information when part of a TA. Unidirectional attestation is insufficient since a module providing data needs assurances regarding the state of the receiver of that data, otherwise it could be sending data to an insecure or malicious module, and the module receiving data needs assurances regarding the state of the sender of the data, otherwise it could be receiving data from an insecure or malicious sender.

In distributed systems there are two, or more, components in different locations communicating through a network. These components may be controlled by a single stakeholder, for example, a banking mobile application communicating with its remote servers, or by multiple stakeholders, for example, multiple chat applications each created by different stakeholders communicating with common remote servers run by a third party. When these components are TAs they must authenticate each other, without trusting the traditional execution environment, before exchanging information. While this could be achieved by running a RA protocol in each direction it is beneficial to have mutual attestation. Distributed systems are frequently located across heterogeneous platforms each supporting a different type of TEE for which reason cross-TEE MA is of particular importance.

\subsubsection*{Contributions}

Our main contributions are a generic protocol for cross-TEE MA (\sref{sec:hema}) and its formal verification (\sref{sec:hema:proofs}).

\subsubsection*{Outline}

\autoref{sec:background} presents background on Tamarin, the prover used in our security analysis of the protocol, and on RA.
\autoref{sec:hema} describes the \hema architecture and protocol.
\autoref{sec:hema:policies} discusses how trust is achieved.
\autoref{sec:hema:proofs} presents our formal model of the \hema protocol and the security properties that we proved.
\autoref{sec:hema:related} presents related work and \autoref{sec:hema:end} concludes this paper.

\section{Preliminaries}
\label{sec:background}

\subsection{Tamarin Prover}
\label{sec:tamarin}

The Tamarin prover~\cite{Staub:BSc:11, Tamarin:manual:2405, Basin:SIGLOG:17, Cortier:ESORICS:20} is a tool for modeling and analyzing security protocols~\cite{Cremers:SEC:20}. The modeler creates a specification of the security protocol and writes the desired security properties. The tool receives the protocol model and the security properties as input, and outputs a proof of correctness for each security property, or a counterexample. In some instances the tool does not terminate because the verification of security protocols in general is undecidable, and in other instances it may \emph{appear} to not terminate because insufficient time was allocated to run the proof. The latter problem happens when $x$ time units are allocated but the tool requires $x+1$ to terminate.

The protocol model specifies the actions taken by the agents, that is, the protocol participants running the protocol, using a language based on multiset rewriting rules, terms and facts. These rules define a transition from one protocol state to the next. Each state contains a symbolic representation of the agents' and the adversary's knowledge, and messages on the network. The agents and the adversary move the protocol to its next state by generating new messages and updating the current ones on the network. Messages are modeled as terms over which functions, whether built-in or user-defined, can be applied. Facts model the generation of fresh values, the interaction with the network, and agent state. Action facts are a special kind of facts that do not appear in the protocol state, only on its trace, and are used by security properties to reason about the protocol's behavior.

The security properties, called lemmas in Tamarin, are modeled using first-order logic and action facts, and proved, or disproved, using either an automated mode or an interactive mode. In the automated mode the proofs are guided by heuristics and in the interactive mode the modeler manually explores and guides the proof.

\subsection{Remote Attestation}
\label{sec:ra}

In RA, a Relying Party validates the security state of an Attester with the assistance of a Verifier. The \emph{Relying Party} wants to ensure the client it is talking to is running on a real TEE, and that both the TEE and the TA image are up to date. The \emph{Attester} wants to prove its state to the Relying Party. The \emph{Verifier} appraises the Attester's Attestation Evidence and produces an Attestion Report as a result. The \emph{Attestation Evidence} contains claims about the Attester, for example, the version of its TEE or a measurement of the instantiated TA image. The \emph{Attestation Report} contains claims made by the Verifier about what it could infer from the Attester taking into account the Attestation Evidence and other information it has access to, for example, reference values with which to compare the contents of the the Attestation Evidence. The Attestation Evidence is received directly from the Attester or routed through the Relying Party, depending on the topological pattern in use (see \aref{sec:topologies}). In some cases, the roles of Relying Party and Verifier may be combined.
In \aref{sec:hera:more} we discuss the minimum requirements for the Attestation Evidence and the Attestation Report, and detail how a protocol for RA might look like.

The MA protocol, presented in the next section, uses a bare-bones RA protocol underneath to obtain Attestation Reports with which the two Attesters authenticate each other.
We define a primitive, \lstinline|ra|, to abstract how that protocol looks like since it depends on the TEE hosting the TA. This primitive receives an optional message as input and outputs an Attestation Report containing the security state of its caller. The optional message is typically used to authenticate some information exchanged between the peers, for example, a public key.

\section{Heterogeneous Mutual Attestation}
\label{sec:hema}

This section presents the \hema protocol, possible topological patterns for the organization of its architecture, how Attesters decide which Verifiers to trust for their own attestation and for the attestation of their peers, that is, other Attesters, and how Attesters decide which peers it is secure to interact with.

\subsection{Threat Model}
\label{sec:hema:tm}

The adversary controls all unprivileged and privileged software on the system, including operating systems and hypervisors. The adversary controls the public network, that is, they can read, modify, delay, and discard any message and inject new messages into the network. Denial of service, rollback~\cite{Brandenburger:DSN:17, Matetic:SEC:17}, and side-channel~\cite{Bulck:SEC:18, Chen:EuroSP:19, Murdock:SP:20, Alder:ACSAC:20} attacks are out of scope. These are orthogonal problems and have their own solutions~\cite{Ahmad:NDSS:18, Oleksenko:ATC:18, Brasser:ACSAC:19}. The TEE is implemented and operating correctly, and its private and secret keying material is not compromised.

The TCB consists of TA instances, for example, enclaved applications in SGX and trusted applications in TrustZone, and the hardware and software required to operate the TEE securely. This hardware and software usually includes the processor package and its microcode and may include a trusted operating system~\cite{Jauernig:M:SP:20}.

\subsection{Architecture}

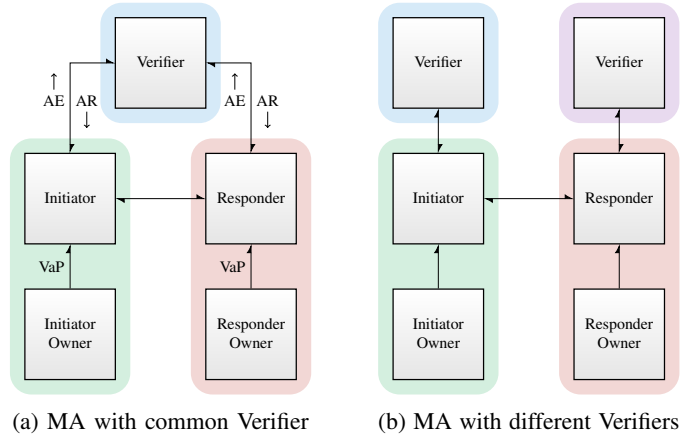
\begin{figure}
	\centering
	\begin{subfigure}[t]{0.45\linewidth}
		\centering
		\begin{tikzpicture}[
	node distance=1cm,
	arrow1/.style={
		-{Latex[scale width=1.5,harpoon]},
	},
	arrow2/.style={
		{Latex[scale width=1.5,harpoon]}-{Latex[scale width=1.5,harpoon]},
	},
	role3/.style={
		draw, rectangle,
		minimum size=2cm,
		align=center, text depth=0.25ex,
		top color=white, bottom color=black!10,
		thick,
	},
	role4/.style={
		draw, rectangle,
		minimum width=2cm, minimum height=1cm,
		align=center, text depth=0.25ex,
		top color=white, bottom color=black!10,
		thick,
	},
	every fit/.style={inner sep=3ex},
]
	\node [role3] (V) at (2,3) {Verifier};
	\node [role3] (I) at (0,0) {Initiator};
	\node [role3] (R) at (4,0) {Responder};
	\node [role3] (IOwner) at (0,-3) {Initiator\\Owner};
	\node [role3] (ROwner) at (4,-3) {Responder\\Owner};

	\draw [arrow2] (I) |- node (label-E5) [left, pos=0.29] {AE} node (label-R6) [right, pos=0.29] {AR} (V);
	\draw [arrow2] (R) |- node (label-E2) [left, pos=0.29] {AE} node (label-R3) [right, pos=0.29] {AR}  (V);
	\draw [arrow2] (I) -- (R);
	\draw [arrow1] (IOwner) -- node [left] {VaP} (I);
	\draw [arrow1] (ROwner) -- node [left] {VaP} (R);

	\coordinate (coordinate-E5) at ($(label-E5)+(0,0.6)$);
	\draw [-Classical TikZ Rightarrow] (label-E5) -- (coordinate-E5);

	\coordinate (coordinate-R6) at ($(label-R6)+(0,-0.6)$);
	\draw [-Classical TikZ Rightarrow] (label-R6) -- (coordinate-R6);

	\coordinate (coordinate-E2) at ($(label-E2)+(0,0.6)$);
	\draw [-Classical TikZ Rightarrow] (label-E2) -- (coordinate-E2);

	\coordinate (coordinate-R3) at ($(label-R3)+(0,-0.6)$);
	\draw [-Classical TikZ Rightarrow] (label-R3) -- (coordinate-R3);

	\begin{scope}[on background layer]
		\node [rounded corners=3ex, fill={rgb,255: red,214; green,234; blue,248}, fit=(V), inner sep=2ex] {};
		\node [rounded corners=3ex, fill={rgb,255: red,212; green,239; blue,223}, fit=(I)(IOwner), inner sep=2ex] {};
		\node [rounded corners=3ex, fill={rgb,255: red,242; green,215; blue,213}, fit=(R)(ROwner), inner sep=2ex] {};
	\end{scope}
\end{tikzpicture}
		\caption{MA with common Verifier}
		\label{fig:arq:ma:1}
	\end{subfigure}
	\hfill
	\begin{subfigure}[t]{0.45\linewidth}
		\centering
		\begin{tikzpicture}[
	node distance=1cm,
	arrow1/.style={
		-{Latex[scale width=1.5,harpoon]},
	},
	arrow2/.style={
		{Latex[scale width=1.5,harpoon]}-{Latex[scale width=1.5,harpoon]},
	},
	role3/.style={
		draw, rectangle,
		minimum size=2cm,
		align=center, text depth=0.25ex,
		top color=white, bottom color=black!10,
		thick,
	},
	role4/.style={
		draw, rectangle,
		minimum width=2cm, minimum height=1cm,
		align=center, text depth=0.25ex,
		top color=white, bottom color=black!10,
		thick,
	},
	every fit/.style={inner sep=3ex},
]
	\node [role3] (V1) at (0,3) {Verifier};
	\node [role3] (V2) at (4,3) {Verifier};
	\node [role3] (I) at (0,0) {Initiator};
	\node [role3] (R) at (4,0) {Responder};
	\node [role3] (IOwner) at (0,-3) {Initiator\\Owner};
	\node [role3] (ROwner) at (4,-3) {Responder\\Owner};

	\draw [arrow2] (I) -- (V1);
	\draw [arrow2] (R) -- (V2);
	\draw [arrow2] (I) -- (R);
	\draw [arrow1] (IOwner) -- (I);
	\draw [arrow1] (ROwner) -- (R);

	\begin{scope}[on background layer]
		\node [rounded corners=3ex, fill={rgb,255: red,214; green,234; blue,248}, fit=(V1), inner sep=2ex] {};
		\node [rounded corners=3ex, fill={rgb,255: red,232; green,218; blue,239}, fit=(V2), inner sep=2ex] {};
		\node [rounded corners=3ex, fill={rgb,255: red,212; green,239; blue,223}, fit=(I)(IOwner), inner sep=2ex] {};
		\node [rounded corners=3ex, fill={rgb,255: red,242; green,215; blue,213}, fit=(R)(ROwner), inner sep=2ex] {};
	\end{scope}
\end{tikzpicture}
		\caption{MA with different Verifiers}
		\label{fig:arq:ma:2}
	\end{subfigure}
	\caption{The architecture and conceptual data flow.}
	\label{fig:arq:ma}
\end{figure}

\autoref{fig:arq:ma:1} and \autoref{fig:arq:ma:2} show a generic architecture for MA and the data flow between its different roles. In comparison to RA, the Attester and the Relying Party are replaced by an Initiator and a Responder. In addition, while in RA there is one Verifier since only one of the parties is being attested, in MA there may be one or two Verifiers depending on whether the Initiator and the Responder use a common Verifier (\autoref{fig:arq:ma:1}) or different Verifiers (\autoref{fig:arq:ma:2}).

\subsubsection{Roles}
\label{sec:hema:roles}

There are three main roles and two support roles, all of which are trusted. The main roles are the Initiator, the Responder, and the Verifier. The support roles are the Initiator Owner and the Responder Owner. The Initiator and the Responder are running on TEEs which can be of the same or different types. The Verifier is located within a trusted environment which may be a hardware-backed TEE but could also be only a trusted server.

\paragraph{Initiator}

The Initiator wants to mutually attest with the Responder and requests that a communication channel be established between both parties. The Initiator receives the Responder's Attestation Report, which it uses to validate the Responder's security state, and in return sends its own Attestation Report, obtained from a trusted Verifier, to the Responder. At the end of a successful protocol run, the Initiator and the Responder have each other's public key from which they can derive shared session keys.

\paragraph{Responder}

The Responder wants to mutually attest with the Initiator and listens to communication requests from possible Initiators. The Responder sends its Attestation Report, obtained from a trusted Verifier, to the Initiator and in returns receives the Initiator's Attestation Report which it uses to validate the Initiator's security state. At the end of a successful protocol run, the Initiator and the Responder have each other's public key from which they can derive shared session keys.

\paragraph{Initiator Owner}

The Initiator Owner provides the Validation Policy with which the Initiator appraises the Responder's Attestation Report.

\paragraph{Responder Owner}

The Responder Owner provides the Validation Policy with which the Responder appraises the Initiator's Attestation Report.

\subsection{Functions}
\label{sec:hema:fun}

\autoref{tab:hema:fun} shows the functions used by the \hema protocol which is presented in the next section (\sref{sec:hema:protocol}).

\begin{table*}
	\centering
	\caption{Functions used by the \hema protocol.}
	\label{tab:hema:fun}
	\begin{tabulary}{\textwidth}{@{}llL@{}}
		\toprule
		Caller & Function & Description \\
		\midrule

		Attester, RP, Verifier & $\rfun{fresh}{}{x}$ & Generates a fresh value $x$ where $x$ can be a nonce $N$, a secret key $K$, or a key pair $(d,Q)$. \\

		Attester & $\rfun{ra}{m}{\var{AR}}$ & The RA primitive runs the RA protocol between the Attester and the Verifier with an optional message $m$ that is included in the $\var{AE}$ as well as in the $\var{AR}$ since $\var{AE}$ is itself included in $\var{AR}$. \\

		RP, Verifier & $\rfun{getpol}{}{\var{P}}$ & Retrieves a policy $P$ where $P$ can be a Validation Policy, $\var{VaP}$, in the RP or a Verification Policy, $\var{VeP}$, in the Verifier. \\

		RP & $\rfun{validate}{\var{AR}, m, \var{VaP}}{\var{state}}$ & Appraises an Attestation Report, $\var{AR}$, based on the Validation Policy, $\var{VaP}$, and the optional bitstring $m$. Returns 1 if the Attester is trusted and 0 otherwise. \\

		Initiator & $\rfun{ma}{}{(d_\var{self}, Q_\var{self}), Q_\var{peer}}$ & Run the mutual attestation protocol. The resulting ephemeral keys -- own private key and peer public key -- can be used to derive session keys with ECDH and a KDF to establish a secure communication channel. \\

		Initiator, Responder & $\rfun{hash}{m}{h}$ & Computes a cryptographic hash function over message $m$. \\

		\bottomrule
	\end{tabulary}
\end{table*}

\subsection{\hema Protocol}
\label{sec:hema:protocol}

This section describes the \hema (Heterogeneous Remote Attestation) protocol. The \hema protocol enables cross-TEE MA.

\begin{figure}
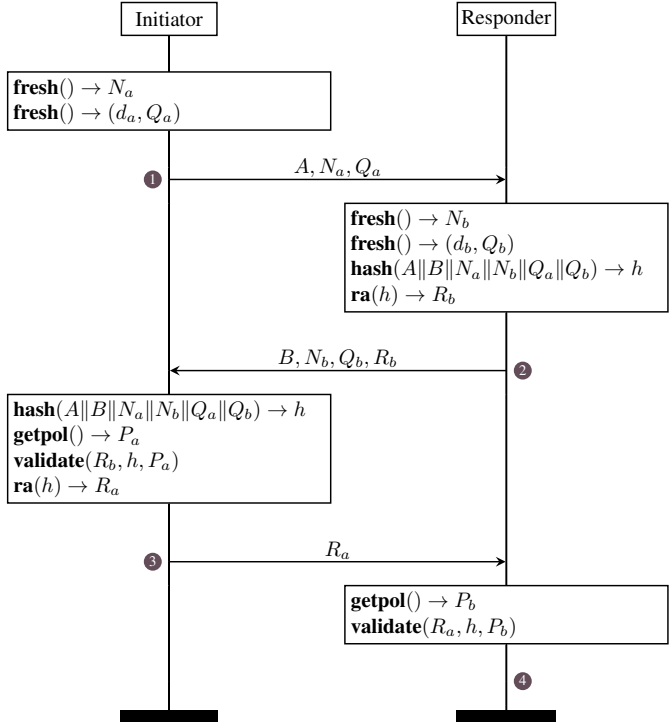

	\centering
	\includestandalone[width=\linewidth]{sd-hema-0}
	\caption[The \hema protocol]{The \hema protocol. Two peers, an Initiator and a Responder, attest each other and exchange ephemeral public keys.}
	\label{fig:hema:sd:0}
\end{figure}

\autoref{fig:hema:sd:0} shows the sequence diagram for the \hema protocol between an Initiator and a Responder.
The protocol steps are the following:

\begin{enumerate}

	\item The Initiator generates a fresh nonce, $N_a$, and a fresh key pair, $(d_a,Q_a)$, and sends its identity, $A$, the nonce $N_a$ and the public key $Q_a$ to the Responder.

	\item The Responder generates a fresh nonce, $N_b$, and a fresh key pair, $(d_b,Q_b)$. Then, it obtains an Attestation Report, $R_b$, certified by a Verifier for its type of TEE using the \hera protocol. As input to the function \fun|ra| of the \hera protocol, the Responder uses a hash, $h$, computed over the identities, nonces, and public keys of both Initiator and Responder. Finally, the Responder replies to the Initiator with its identity, $B$, the nonce $N_b$, the public key $Q_b$, and the report $R_b$.

	\item The Initiator computes a hash value, $h$, over the identities, nonces, and public keys of both Initiator and Responder. Then, the Initiator validates the Attestation Report, $R_b$, received from the Responder. In particular, it
	(i) verifies the signature over $R_b$ to make sure it is correctly signed by a trusted Verifier,
	(ii) checks whether the TCB level field (see \sref{sec:hera:obj}) on the report states that the Responder is a legitimate and up-to-date TA instance running on a legitimate and up-to-date TEE, and
	(iii) compares the hash it previously computed with the one stored in $R_b$ to ensure these match which implies the public keys have been correctly exchanged and the Attestation Report $R_b$ received from the Responder is fresh.
	Finally, the Initiator obtains a report, $R_a$, from a Verifier for its type of TEE using the \hera protocol with $h$ as input to the function \fun{ra}, and sends $R_a$ to the Responder.

	\item The Responder validates the Attestation Report, $R_a$, received from the Initiator similarly to how the Initiator validated the Attestation Report $R_b$.

\end{enumerate}

At the end of a successful protocol run, the Initiator and the Responder have their own ephemeral key pair and their peer's ephemeral public key. These keys can be used, for example, to create a trusted communication channel between one another.

\section{Attestation Policies}
\label{sec:hema:policies}

Attestation informs an entity about the state of a particular system with a high degree of confidence~\cite{Martin:TR:08:IntroTC}. However, even if the system is considered to be in a good state, that is, the software and hardware are legitimate and up to date, it does not imply it can be trusted. Nevertheless, this knowledge helps said entity making a decision on whether to trust the Attester, and what level of trust should be granted in case there is more than a \emph{full trust} or \emph{zero trust} binary choice. This entity is the Relying Party in RA (see \sref{sec:ra}), and the Initiator and the Responder in MA (see \sref{sec:hema}).

\subsection{Trusting Attesters}
\label{sec:hema:policies:attester}

In RA, the Relying Party evaluates the trustworthiness of an Attester and decides whether to trust that Attester by applying a Validation Policy to the Attestation Report. The Validation Policy is provisioned by the Relying Party Owner and the Attestation Report is produced by a Verifier.

The Validation Policy is logically separate from the TA image and signed by the same private key that signs the TA image, or by a different private key that can be hardcoded into the TA image, in order to bind the Validation Policy to a TA image. The Validation Policy contains the requirements for trusting a target environment, that is, the characteristics that the Relying Party performing the appraisal deems acceptable or not acceptable. Having the Validation Policy hardcoded into the TA causes problems if it contains TA image measurements because by adding such information to the TA code it also changes its measurement. This is similar to a known problem where two TAs want to hardcode each other's identities but by doing so change their own identity which is typically a cryptographic hash of the TA code, data, and initial configuration~\cite{Beekman:AsiaCCS:16}.

The Verifier produces the Attestation Report by applying a Verification Policy to the Attestation Evidence it receives from the Attester. The Verification Policy is provisioned by those in control of the Verifier. In addition, the Verifier may have access to reference values other than what is already part of the Verification Policy, and knows it can trust the Attestation Evidence because the Attester is in possession of a root of trust for reporting~\cite{Anati:HASP:13}.

In MA there is a similar process, but instead of a Relying Party there is an Initiator and a Responder each of which takes the roles of both Attester and Relying Party. The Initiator provides evidence about itself to prove to the Responder it is trustworthy while simultaneously evaluating the Responder's trustworthiness and appraising the Responder's Attestation Report, and vice versa.

\subsection{Trusting Verifiers}
\label{sec:hema:policies:verifier}

During RA the Relying Party, which in the case of \hema is another Attester, called Initiator or Responder depending on its role, receives an Attestation Report containing information on the state of the Attester the report pertains to. This Attestation Report is signed by the Verifier which is a trusted party. Considering anyone can sign a fake Attestation Report this raises the question \emph{how does the Relying Party know which Verifier(s) to trust?}

In the simple case there is a system with a single TEE type and one Verifier. This is the case in SGX when used with the IAS (Intel Attestation Service)~\cite{Intel:20:IASr6}. In this situation, where the Verifier is one well-known service used by everyone, the protocol participants can hardcode the Verifier's long-term public key to ensure the Attestation Reports are legitimate.

In a system with cross-TEE attestation and multiple Verifiers, hardcoding the Verifiers' long-term public keys is still a viable, and the simplest, solution. However, since the system is now dealing with multiple TEE types and Verifiers, having all stakeholders agree on the same set of Verifiers to trust is not a simple task.

This section discusses alternative solutions to hardcoding the Verifiers' public keys, taking into account in whom trust is placed. We make the assumption that each Verifier is represented by a public key. The Verifier could sign Attestation Reports using the corresponding private key, or more likely, have a long-term key pair where the long-term public key is delivered to its clients and the long-term private key certifies the key pair actually signing the Attestation Reports.

Each of the four proposed solutions has its own advantages and disadvantages. These solutions differentiate themselves based on who issues the list of trusted verifiers and how that list is distributed.
The list of trusted verifiers can be:
\begin{enumerate}
	\item hardcoded in the TA image;

	\item part of a local configuration file;

	\item provided by a web service; or

	\item contained in the Attestation Report.
\end{enumerate}

\paragraph{Hardcoded in the TA image}

The list of trusted verifiers is issued by the ISV and hardcoded in the TA image. In this case, the TA instances verify whether the entity signing the Attestation Reports they receive is part of this list and if not they reject the report.
Adding or removing a Verifier from the trusted list requires updating the TA image.

\paragraph{Part of a local configuration file}

The trusted verifiers list can be part of a local configuration file loaded by TA instances. The local file is signed by a private key with the corresponding public key hardcoded in the TA image. The issuer of the file containing the list owns this key pair and can be the ISV or another trusted third party.
An advantage of this solution over the previous one is it decouples the list of trusted verifiers from the TA image enabling independent updates. However, there is a problem of freshness. \emph{How does a TA instance knows the file with the list of trusted verifiers is the latest one?} In the case of the TA image we can assume the TEE runtime only accepts instances created from the most recent version of a TA image, otherwise remote and mutual attestation fail. In the case of the configuration file, if a new configuration file is issued the TA instance cannot distinguish between the current version and a previous version of the file. A possible solution is hardcoding a common nonce in both TA image and configuration file in essence binding one to the other. But now an update to the configuration file requires updating the TA image removing the decoupling advantage of this solution.

\paragraph{Provided by a web service}

A web service can provide the list of trusted verifiers on request to TA instances. The system can have multiple web services and each ISV decides in which one(s) to place their trust. Any entity can issue and distribute the list of trusted verifiers through this medium. For example, the ISV could run a web service only for its own TAs, a TEE provider could offer this service to its clients, a consortium of TEE providers could come together to offer this service to their clients, or the entity behind a Verifier could provide a distribution service in addition to attestation services.
Adding or removing entities from the list of trusted verifiers is done by updating the list on the web service side. The effects can propagate quickly depending on how often the TEE instances obtain a copy of the list. A TA instance could obtain a copy of the list from the web service when the instance is created, at a minimum, unless it prefers to seal the existing list. Sealing would trade a request to a web service by having to maintain the list in disk which has the downside that its freshness cannot be ensured -- not without making a request to the web service in which case the TA instance is better off simply requesting the new up-to-date list of trusted verifiers.

\paragraph{Contained in the Attestation Report}

Attestation Reports are already produced and signed by Verifiers during attestation. Adding a list of trusted verifiers to the Attestation Report allows each Verifier to decide which other Verifiers it trusts. Such trust could be done by agreement, for example, Arm and Intel could each have a Verifier and agree to trust each other.
A TA instance receives its own and its peer's Attestation Reports during MA. In this case, the TA instance decides which Verifiers to trust based on the list of trusted verifiers in its own Attestation Report, or the union or intersection of the list of trusted verifiers of both Attestation Reports.
Adding or removing entities from the list of trusted verifiers is done by updating it on the Verifier's side. The effects propagate quickly since Attestation Reports are issued during mutual attestation.

\paragraph{Hybrid solution}

Hybrid solutions use multiple sources for the list and can create a set of trusted verifiers by merging or intersecting the lists of different sources. For example, an ISV could hardcode in their TAs its own list of trusted verifiers and still obtain a list from a web service. This could be done to increase redundancy since if the web service is down the hardcoded list is still available, or to have the input of different parties to the set of trusted verifiers by intersecting the hardcoded list issued by the ISV and the web service list issued by a different party.

\paragraph{Comparison}

Each option places control of the list of trusted verifiers in different stakeholders. When hardcoded the ISV issues the list, when part of the configuration file or distributed by a web service the ISV or any other trusted entity can issue the list, and when the list is part of the Attestation Report it is issued by the same stakeholder who controls the Verifier.
%
Updating the list is easier when distributed via web service or added to Attestation Reports than when hardcoded to TA images or part of configuration files which would require redeploying the TA image and/or the configuration file. However, it is fair to assume that such lists are rarely updated as long as Verifiers keep their long-term key pairs secure.
An additional consideration might be on whether the different means to obtain the list of trusted verifiers increase the software stack of the TA image. For the hardcoded solution no particular software is needed; for the configuration file solution the TA instance needs read access to the file system storing the list; and for the web service and Attestation Report solutions the means to access those, possibly remote, services is necessary but this already exists since Attesters and/or Relying Parties already have to contact one such service -- the Verifier(s) -- during mutual attestation.

\section{Cross-TEE MA Security Analysis}
\label{sec:hema:proofs}

This section formalizes the \hema protocol, shown in \autoref{fig:hema:sd:0}, using the Tamarin Prover introduced in \autoref{sec:tamarin}.
The first part (\sref{sec:hema:formal:model}) discusses the model including its setup phase, the actual protocol rules where messages are exchanged between agents using the public network, the assumptions made, and how the model was validated. The second part (\sref{sec:hema:formal:analysis}) describes our proofs and what can be inferred from these results.

\subsection{Model}
\label{sec:hema:formal:model}

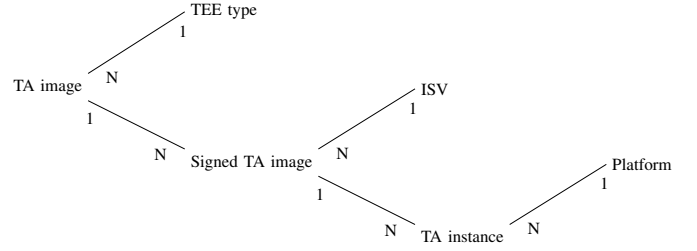
\begin{figure}
	\centering
	\begin{tikzpicture}[%
	entity/.style={},
]
	\node[entity] (instance) {TA instance};
	\node[entity] (signed) [above left=1 and 2 of instance] {Signed TA image};
	\node[entity] (image) [above left=1 and 2 of signed] {TA image};
	\node[entity] (type) [above right=1 and 2 of image] {TEE type};
	\node[entity] (h) [above right=1 and 2 of signed] {ISV};
	\node[entity] (p) [above right=1 and 2 of instance] {Platform};

	\draw (type.west) -- (image.north east)
		node[anchor=north west, very near start] {1}
		node[anchor=north west, very near end] {N};

	\draw (h.west) -- (signed.north east)
		node[anchor=north west, very near start] {1}
		node[anchor=north west, very near end] {N};

	\draw (p.west) -- (instance.north east)
		node[anchor=north west, very near start] {1}
		node[anchor=north west, very near end] {N};

	\draw (image.south east) -- (signed.north west)
		node[anchor=north east, very near start] {1}
		node[anchor=north east, very near end] {N};

	\draw (signed.south east) -- (instance.north west)
		node[anchor=north east, very near start] {1}
		node[anchor=north east, very near end] {N};
\end{tikzpicture}
	\caption{Formal model's objects relationship.}
	\label{fig:hema:er}
\end{figure}

\begin{figure}
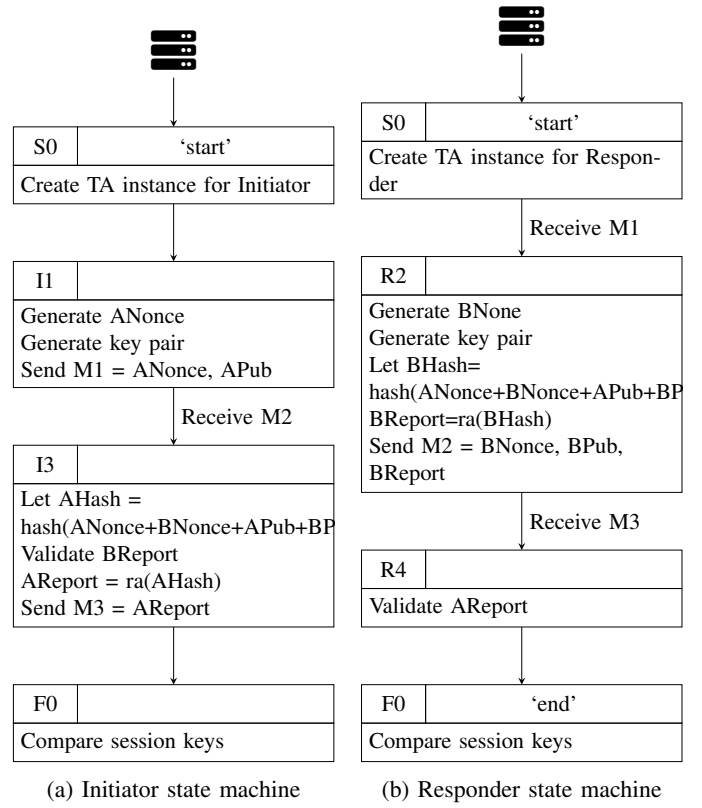

	\centering
	\begin{subfigure}[t]{0.48\linewidth}
		\centering
		\includestandalone[width=\linewidth]{sm1}
		\caption{Initiator state machine}
		\label{fig:hema:sm1}
	\end{subfigure}
	\hfill
	\begin{subfigure}[t]{0.48\linewidth}
		\centering
		\includestandalone[width=\linewidth]{sm2}
		\caption{Responder state machine}
		\label{fig:hema:sm2}
	\end{subfigure}
	\caption[Initiator and Responder state machines]{Initiator and Responder state machines. In both cases, there is a setup state, S0, that creates an instance for the agent, and a final state, F0, that compares the agents session keys for equality. S0 and F0 are not part of the protocol proper in \autoref{fig:hema:sd:0} and do not exchange any messages via the public network.}
	\label{fig:hema:sm}
\end{figure}

\autoref{fig:hema:er} shows the relationship between the objects of the formal model. Each object corresponds to one rule in the prover and together form the setup phase of the model which is depicted as state S0 in \autoref{fig:hema:sm}. The objective is having, at the end of the setup phase, a TA instance running on a platform. This TA instance is initialized with a TA image that is signed by a particular ISV and is of some TEE type. The TA instance is then assigned to an agent by the prover when running proofs.

\begin{itemize}

	\item A TA image has one TEE type. Two TAs designed to provide the exact same service but targeting two different TEEs are unlikely to have the same code because the TEEs' APIs are different and the programming language itself might be different. Since a TA image is typically represented by a measurement over its code and initial data then these two hypothetical TA images would have different measurements and be considered different from one another.


	There could be edge cases where this is untrue. Consider two different TEEs, each with its own underlying code base but adopting the same, standardized, API. An ISV could develop a single TA to run on both TEEs by taking advantage of their shared API. Even in this scenario, each TA would likely have different configurations and sets of initial data, for example, different hardcoded public keys or URLs, which would result in different measurements. In addition, because the underlying code of the TEEs is different, the libraries pulled into the TA during compilation are also different resulting again in different measurements of the TA. For this reason, the assumption that one TA image has a single TEE type makes sense.

	\item A signed TA image is a TA image that has been signed by a private key belonging to the entity owning the TA. The developer of the TA and its owner could be different, for example, when the TA owner delegates development to a third party. For simplicity, and since it does not affter the model, we consider the TA owner and the TA developer to be the same entity.

	The same TA image could be signed by different entities. Consider a scenario where a reference TA, for example, a simple key-value store database, is made available and adopted by different TA owners. Each TA owner could modify setup data for the key-value store, in which case the TA image would become different, or use the key-value store TA as-is in which case the TA image is the same by signed by different entities.

	Note the signature can be one of the characteristics of a signed TA image used to authenticate that image.

	\item A TA owner can sign multiple TA images.

	\item A signed TA image can create any number of TA instances.

	\item A platform can run any number of TA instances.

\end{itemize}

\autoref{fig:hema:sm} shows the state machines for the Initiator and for the Responder. There are two states, S0 and F0, common to both agents. State S0 is a setup state and models the relationships between entities and objects in the model according to \autoref{fig:hema:er}. In practice, S0 is composed of multiple rules but is shown as a single state for simplification. State F0 verifies whether the public key exchange was successful using an equality restriction. This verification could have been done in each lemma but by doing it directly in a common rule at the end of each run it reduces boilerplate code.

S0 and F0 do not send/receive messages to/from the public network. These two states exchange only state facts which the adversary cannot read. From this perspective, S0 and F0 give no advantage to the adversary.

Our model makes a set of assumptions in order to simplify the proofs while simultaneously including the important parts of the \hema protocol.

\subsubsection{Assumptions}

The \hema protocol delegates RA to the layer below which is invoked using $\rfun{ra}{m}{}$. In the model, the Attestation Evidence includes the optional bitstring $m$ containing data passed by the \hema layer but not other claims. This decision simplifies the model without adverse affects because the appraisal of the Attestation Evidence is a matter of policy.

The model assumes the correctness of the underlying RA protocol.

Attestation Evidence and Attestation Reports are produced inside rules R2 and I3 instead of using extra rules to model the Verifier.

The long-term keys of Endorsers and Verifiers meant to sign Attestation Evidence and Attestation Reports, respectively, are replaced by a private function \lstinline|ersk| which stands for evidence--report signing key. The \lstinline|ersk| is used directly in places where the long-term keys, or more usually a derivation based on those long-term keys, would be used. This simplifies the model but also makes it stronger since the compromise of \lstinline|ersk| would imply the compromise of all keys it replaces.

\subsubsection{Modifications to the Model}
\label{sec:hema:formal:model:modifications}

The model follows the protocol presented in \autoref{sec:hema}. A preliminary version of the protocol did not exchange the identities of the Initiator and the Responder, and was subject to a MitM attack as shown in \autoref{lst:hema:MitM}. The problem was that the identities, $A$ and $B$, were not part of the input to the cryptographic hash, that is, instead of $\rfun{hash}{A \| B \| N_{a} \| N_{b} \| Q_{a} \| Q_{b}}{h}$ we had $\rfun{hash}{N_{a} \| N_{b} \| Q_{a} \| Q_{b}}{h}$ which allowed an adversay $\textcolor{red}{C}$ to independently establish connections to both $A$ and $B$.

\begin{lstlisting}[
	language=hema,
	style=protocolo,
	caption={[MitM attack on a preliminary version of the \hema protocol]MitM attack on a preliminary version of the \hema protocol. The identities of the Initiator and the Responder, $A$ and $B$, respectively, are not exchanged in this version of the mutual attestation scheme but are still shown, crossed out, for completeness.},
	label=lst:hema:MitM,
	morekeywords={hash},
]
A->%\textcolor{red}{C}%: $\cancel{A}, N_{a}, Q_{a}$
%\textcolor{red}{C}%->B: $N_{c}, Q_{c}$
B->%\textcolor{red}{C}%: $\cancel{B}, N_{b}, Q_{b}, R_{b}$
%\textcolor{red}{C}%->A: $N_{c}, Q_{c}, R_{c}$
A->%\textcolor{red}{C}%: $R_{a}$
%\textcolor{red}{C}%->B: $R_{c}$
\end{lstlisting}

\subsubsection{Successful Execution}
\label{sec:hema:formal:model:sanity}

Multiple \lstinline|exists-trace| lemmas test whether our model can run to completion successfully. These lemmas are not mandatory but are useful in finding bugs in the formalization of the \hema scheme. If a model cannot complete successfully not only is it wrong but also could result in proofs holding falsely.
For example, we maintain the agents' state in between each of their rules using state facts, which in our model are called \lstinline|AgSt|. In the first rule of the Initiator, I1, the agent stores its identity, nonce, private key, and other variables in a state fact as \lstinline|AgSt(..., <$I, ~ni, ~ki, ...>)| which is later retrieved in the second rule of the Initiator, I3. If we were to forget to include the identity, \lstinline|$I|, in the input to I3, Tamarin would successfully verify the secrecy and agreement security properties, which are discussed next in \autoref{sec:hema:formal:analysis}, even though the model is incorrect. This particular error would have been caught by our \lstinline|exists-trace| lemmas (see \aref{app:hema:miiisf}).

\autoref{tab:hera:formal:exists} shows the sanity checks applied to our model, most of which attempt to complete an execution of the protocol between two agents. There are multiple lemmas so that different combinations of parameters can be tested resulting in a model less likely to contain errors. This is also useful for the future since it increases the likelihood bugs are caught earlier on in case the model is tweaked. The parameters are whether the two agents use the same or different Verifiers (V) or the same or different types of TEE (T), whether the agents are located on the same or different platforms (P), and whether the two agents have the same or different Author MRs (A) or the same or different Image MRs (I).

\begin{table}
	\centering
	\caption[The sanity checks applied to our formal model]{The sanity checks applied to our formal model.}
	\label{tab:hera:formal:exists}
	\begin{tabulary}{\linewidth}{@{}LCCCCCL@{}}
		\toprule
		Lemma & V & T & P & A & I & Description \\
		\midrule
		\lstinline|final_1| & \faIcon{not-equal} & \faIcon{equals} & \faIcon{not-equal} & \faIcon{equals} & \faIcon{not-equal} & Two agents mutually attest themselves successfully. \\
		\lstinline|final_2| & \faIcon{not-equal} & \faIcon{equals} & \faIcon{not-equal} & \faIcon{not-equal} & \faIcon{not-equal} & Two agents mutually attest themselves successfully. \\
		\lstinline|final_3| & \faIcon{not-equal} & \faIcon{not-equal} & \faIcon{not-equal} & \faIcon{equals} & \faIcon{not-equal} & Two agents mutually attest themselves successfully. \\
		\lstinline|final_4| & \faIcon{not-equal} & \faIcon{not-equal} & \faIcon{not-equal} & \faIcon{not-equal} & \faIcon{not-equal} & Two agents mutually attest themselves successfully. \\
		\lstinline|final_5| & \faIcon{not-equal} & \faIcon{equals} & \faIcon{not-equal} & \faIcon{equals} & \faIcon{equals} & Two agents mutually attest themselves successfully. \\
		\lstinline|final_6| & \faIcon{not-equal} & \faIcon{equals} & \faIcon{not-equal} & \faIcon{not-equal} & \faIcon{equals} & Two agents mutually attest themselves successfully. \\
		\lstinline|final_7| & \faIcon{not-equal} & \faIcon{not-equal} & \faIcon{not-equal} & \faIcon{equals} & \faIcon{equals} & Different types of TEE implies different Image MRs. \\
		\lstinline|final_8| & \faIcon{equals} & \faIcon{equals} & \faIcon{not-equal} & \faIcon{equals} & \faIcon{equals} & Two agents mutually attest themselves successfully. \\
		\bottomrule
	\end{tabulary}
	\\ \medskip
	\faIcon{equals} same entity \qquad \faIcon{not-equal} different entities
\end{table}

The lemma \lstinline|final_7| is different from the other sanity checks because it does not use an \lstinline|exists-trace| lemma, which looks for a single instance where a lemma applies, but a formula that must hold on all traces. Note this is similar to a proof lemma, which must hold on all traces as well. The reason behind this lemma is that we assume that two TA images created for two different types of TEE always have different Image MRs. Intuitively this makes sense because if we have the same application compiled for two different TEEs, for example, to run in SGX and to run in TrustZone, the compilation result is highly likely to be different leading to the Image MRs to be different as well. Taking this into account, what \lstinline|final_7| states is that it is not possible to have two agents with different types of TEE but with the same Image MR successfully attesting each other simply because there is a reasonable assumption that this does not happen in the real world. Logically, having two different types of TEE implies having two different Image MRs.

\subsection{Analysis}
\label{sec:hema:formal:analysis}

This section discusses our proofs for the secrecy of the session keys and for the agreement security properties.

\subsubsection{Secrecy of the Session Keys}

The secrecy proof is applied to the individual session private keys of Initiator and Responder, and not to the resulting shared key. The rationale is that by the end of the \hema protocol both agents are mutually attested and have each other's session public key. What the Attesters then do with those keys, or how a secure communication channel is set up using those keys, is up to the TA developer.

The lemma states that when an action \lstinline|Secret(x)| occurs at timepoint \lstinline|i|, the adversary does not know \lstinline|x|. The statement is the following:

\begin{lstlisting}
	All x #i. Secret(x) @i ==> not (Ex #j. K(x) @j)
\end{lstlisting}
\smallskip

\subsubsection{Injective Agreement}

The agreement security properties follow the definition given by Lowe in ``a hierarchy of authentication specifications''~\cite{Lowe:CSFW:97}. The idea behind these definitions -- aliveness, weak agreement, non-injective agreement, and injective agreement -- is finding what an agent A can deduce about an agent B after completing an execution of the protocol, apparently with B. In our case, A and B are the Initiator and the Responder, respectively.

The four agreement proofs are written from the perspective of both Initiator and Responder resulting in eight proofs in total. In addition, we do small changes to the model in order to see what the effects are in relation to the protocol version presented in \autoref{sec:hema}. \autoref{tab:hema:formal:results} shows the results of running the proofs in the Tamarin prover. There are three versions of the model: \lstinline|hema| is the original model representing the protocol as described in \autoref{sec:hema}, \lstinline|hema-ni| is a simplified version where the agents neither exchange identities nor include them in the user data field of the Attestation Evidence, and \lstinline|hema-nie| is another simplified version where the agents do not exchange the identities but still add them to the user data field. All the proofs are verified successfully for the \lstinline|hema| and \lstinline|hema-nie| versions of the model, but the agreement properties fail for the \lstinline|hema-ni| version of the model where the identities of the Initiator and the Responder are not included in the user data exchanged during the protocol. The \lstinline|hema-ni| version of the model corresponds to the preliminary version of \lstinline|hema| that is subject to a MitM attack as explained in \ref{sec:hema:formal:model:modifications}.

\begin{table}
	\centering
	\caption{The results of running the Tamarin prover.}
	\label{tab:hema:formal:results}
	\begin{tabulary}{\linewidth}{@{}LTTTTTT@{}}
		\toprule
		& \multicolumn{2}{c}{hema} & \multicolumn{2}{c}{hema-ni} & \multicolumn{2}{c}{hema-nie} \\
		\cmidrule(lr){2-3} \cmidrule(lr){4-5} \cmidrule(lr){6-7}
		Security Property & I & R & I & R & I & R \\
		\midrule
		\textbf{Result} &&&&&& \\
		Secrecy                 &  v  &  v  &  v  &  v  &  v  &  v  \\
		Aliveness               &  v  &  v  &  f  &  f  &  v  &  v  \\
		Weak agreement          &  v  &  v  &  f  &  f  &  v  &  v  \\
		Non-injective agreement &  v  &  v  &  f  &  f  &  v  &  v  \\
		Injective agreement     &  v  &  v  &  f  &  f  &  v  &  v  \\
		\addlinespace
		\textbf{No. of steps} &&&&&& \\
		Secrecy                 &  20 &  20 &  20 &  20 &  20 &  20 \\
		Aliveness               &   5 &   4 &  19 &  20 &  29 &  18 \\
		Weak agreement          &   5 &   4 &  19 &  20 &  29 &  18 \\
		Non-injective agreement &   5 &   4 &  19 &  20 &  29 &  28 \\
		Injective agreement     &   9 &  18 &  19 &  20 &  57 & 156 \\
		\bottomrule
	\end{tabulary}
	\\ \medskip
	\textcolor{pass}{v} -- lemma verified \qquad \textcolor{fail}{f} -- lemma falsified
\end{table}

\section{Related Work}
\label{sec:hema:related}

\begin{figure}
	\centering
	\begin{tikzpicture}[
	node distance=1cm,
	x=1cm, y=1cm, z=0.6cm,
	Classical TikZ Rightarrow-Classical TikZ Rightarrow,
]
	\draw (xyz cs:x=-2) node [left] {intra-platform} -- (xyz cs:x=2) node [right] {inter-platform};
	\draw (xyz cs:y=-2) node [below] {intra-TEE} -- (xyz cs:y=2) node [above] {inter-TEE};
	\draw (xyz cs:z=-2) node [below left, align=center] {unidirectional\\attestation} -- (xyz cs:z=2) node [above right, align=center] {bidirectional\\attestation};

	\coordinate (zero) at (0,0); 
	\coordinate (L) at (-4.2,0);
	\coordinate (R) at (+4.2,0);
	\coordinate (U) at (0,+2.4);
	\coordinate (D) at (0,-2.4);
	\node [fit=(zero)(L)(R)(U)(D)] {};
\end{tikzpicture}
	\caption{Taxonomy of attestation in TEEs.}
	\label{fig:taxonomy}
\end{figure}
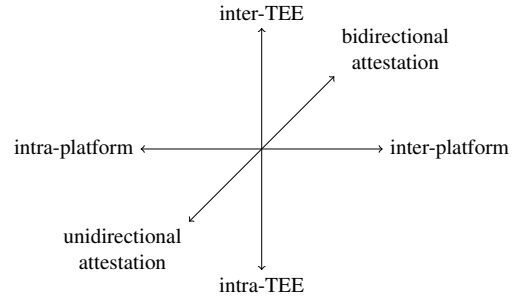

\autoref{fig:taxonomy} shows a taxonomy of attestation in TEEs according to the location of the peers and the attestation direction.
In \emph{intra-platform} attestation the peers are located within the same computer system whereas in \emph{inter-platform} attestation the peers are located in different computer systems.
In \emph{intra-TEE} attestation only one TEE type is supported whereas in \emph{inter-TEE} attestation multiple TEE types are supported.
In \emph{unidirectional} attestation one peer attests another peer whereas in \emph{bidirectional} attestation peers mutually attest each other.
Bidirectional attestation can be achieved by using RA in both directions, or by using a dedicated MA scheme such as \hera.

\subsection{Remote Attestation}

In RA, a Relying Party validates the trustworthiness of an Attester executing inside a TEE. This is the case, for example, in Intel SGX's EPID~\cite{Intel:16:EPID} and DCAP~\cite{Intel:18:DCAP} solutions as well as in OPERA~\cite{Chen:CCS:19}. OPERA also targets SGX, improving enclave privacy, but still relies on EPID to establish a root of trust to the SGX platforms.
\hema is different since it targets bidirectional attestation instead of unidirectional attestation, and works across different TEE types.

In collective remote attestation (CRA)~\cite{Ambrosin:COMST:20}, we typically have unidirectional attestation applied to a swarm of devices by a Verifier with the result being the overall health of the swarm, and optionally, of its individual nodes. \hema could be integrated into the concept of CRA to provide bidirectional attestation.

\subsection{Mutual Attestation}

In MA, two Attesters validate each other's trustworthiness and both are expected to be executing inside TEEs.
MAGE~\cite{Chen:SEC:22} and IPAS~\cite{Andrade:PRDC:23} are MA solutions for SGX that work without trusted third parties, however, MAGE does not support updates to the enclaves.
Teaclave~\cite{Teaclave:22:web:MA} aims to support MA for both SGX and TrustZone but currently only SGX is supported, and its MA attestation solution uses third-party auditors that sign and publish the identities of trusted enclaves.
\hema supports enclave updates and works, without requiring trusted third parties, for both the same TEE type and, in addition, to different TEE types which the previous solutions do not support.

\section{Conclusion}
\label{sec:hema:end}

We proposed an architecture and protocol to mutually attest TA instances executing on the same or on different platforms, and on the same or on different TEE types, which is crucial for building cross-TEE distributed systems.
TAs can use \hema to ensure their peers meet the expected security conditions before exchanging data, even when those peers are running on TAs hosted by a different TEE type.
Our protocol is formally verified using the Tamarin prover.

\section*{Acknowledgments}
\addcontentsline{toc}{section}{Acknowledgments}

This work was financially supported by Project Blockchain.PT -- Decentralize Portugal with Blockchain Agenda (Project no 51), WP 6: Digital Assets Management, Call no 02/C05-i01.01/2022, funded by the Portuguese Recovery and Resilience Program (PRR), The Portuguese Republic, and The European Union (EU) under the framework of the Next Generation EU Program. This work was also supported by national funds through Fundação para a Ciência e a Tecnologia, I.P. (FCT) under projects UID/50021/2025 (DOI: \url{https://doi.org/10.54499/UID/50021/2025}).

\bibliographystyle{IEEEtran}
\bibliography{references}

\begin{thebibliography}{10}
\providecommand{\url}[1]{#1}
\csname url@samestyle\endcsname
\providecommand{\newblock}{\relax}
\providecommand{\bibinfo}[2]{#2}
\providecommand{\BIBentrySTDinterwordspacing}{\spaceskip=0pt\relax}
\providecommand{\BIBentryALTinterwordstretchfactor}{4}
\providecommand{\BIBentryALTinterwordspacing}{\spaceskip=\fontdimen2\font plus
\BIBentryALTinterwordstretchfactor\fontdimen3\font minus
  \fontdimen4\font\relax}
\providecommand{\BIBforeignlanguage}[2]{{%
\expandafter\ifx\csname l@#1\endcsname\relax
\typeout{** WARNING: IEEEtran.bst: No hyphenation pattern has been}%
\typeout{** loaded for the language `#1'. Using the pattern for}%
\typeout{** the default language instead.}%
\else
\language=\csname l@#1\endcsname
\fi
#2}}
\providecommand{\BIBdecl}{\relax}
\BIBdecl

\bibitem{Armbrust:M:CACM:10}
M.~Armbrust, A.~Fox, R.~Griffith, A.~D. Joseph, R.~Katz, A.~Konwinski, G.~Lee,
  D.~Patterson, A.~Rabkin, I.~Stoica, and M.~Zaharia, ``A view of cloud
  computing,'' \emph{Communications of the {ACM}}, vol.~53, no.~4, pp. 50--58,
  Apr. 2010.

\bibitem{Rocha:DCDV:11:6}
F.~Rocha and M.~Correia, ``Lucy in the sky without diamonds: Stealing
  confidential data in the cloud,'' in \emph{Proceedings of the 1st
  International Workshop on Dependability of Clouds, Data Centers and Virtual
  Computing Environments ({DCDV})}, Jun. 2011, pp. 129--134.

\bibitem{Maene:IEEE:TC:18}
P.~Maene, J.~Götzfried, R.~de~Clercq, T.~Müller, F.~Freiling, and
  I.~Verbauwhede, ``Hardware-based trusted computing architectures for
  isolation and attestation,'' \emph{{IEEE} Transactions on Computers ({TC})},
  vol.~67, no.~3, Mar. 2018.

\bibitem{Singh:M:Queue:21}
J.~Singh, J.~Cobbe, D.~L. Quoc, and Z.~Tarkhani, ``Enclaves in the clouds:
  Legal considerations and broader implications,'' \emph{{ACM} Queue}, vol.~18,
  no.~6, pp. 78--114, Jan. 2021.

\bibitem{McKeen:HASP:13}
F.~McKeen, I.~Alexandrovich, A.~Berenzon, C.~Rozas, H.~Shafi, V.~Shanbhogue,
  and U.~Savagaonkar, ``Innovative instructions and software model for isolated
  execution,'' in \emph{Proceedings of the 2nd International Workshop on
  Hardware and Architectural Support for Security and Privacy ({HASP})}, Jun.
  2013.

\bibitem{Arm:09:TrustZone}
\emph{{ARM} Security Technology -- Building a Secure System using {TrustZone}
  Technology}, {ARM} Limited, Apr. 2009.

\bibitem{DoD:85:TCSEC}
\emph{Trusted Computer System Evaluation Criteria}, Department of Defense, Dec.
  1985, doD 5200.28-STD.

\bibitem{Tamarin:manual:2405}
{The Tamarin Team}, \emph{Tamarin-Prover Manual}, May 2024.

\bibitem{Staub:BSc:11}
C.~Staub, ``A user interface for interactive security protocol design,''
  Bachelor Thesis, ETH Zurich, 2011.

\bibitem{Basin:SIGLOG:17}
D.~Basin, C.~Cremers, J.~Dreier, and R.~Sasse, ``Symbolically analyzing
  security protocols using {Tamarin},'' \emph{{ACM} {SIGLOG} News}, vol.~4,
  no.~4, pp. 19--30, Oct. 2017.

\bibitem{Cortier:ESORICS:20}
V.~Cortier, S.~Delaune, and J.~Dreier, ``Automatic generation of sources lemmas
  in {Tamarin}: Towards automatic proofs of security protocols,'' in
  \emph{Proceedings of the 25th European Symposium on Research in Computer
  Security ({ESORICS})}, Sep. 2020.

\bibitem{Cremers:SEC:20}
C.~Cremers, B.~Kiesl, and N.~Medinger, ``A formal analysis of {IEEE} 802.11's
  {WPA2}: Countering the kracks caused by cracking the counters,'' in
  \emph{Proceedings of the 29th {USENIX} Security Symposium}, Aug. 2020.

\bibitem{Brandenburger:DSN:17}
M.~Brandenburger, C.~Cachin, M.~Lorenz, and R.~Kapitza, ``Rollback and forking
  detection for trusted execution environments using lightweight collective
  memory,'' in \emph{Proceedings of the 47th Annual {IEEE/IFIP} International
  Conference on Dependable Systems and Networks ({DSN})}, Jun. 2017, pp.
  157--168.

\bibitem{Matetic:SEC:17}
S.~Matetic, M.~Ahmed, K.~Kostiainen, A.~Dhar, D.~Sommer, A.~Gervais, A.~Juels,
  and S.~Capkun, ``{ROTE}: Rollback protection for trusted execution,'' in
  \emph{Proceedings of the 26th {USENIX} Security Symposium}, Aug. 2017.

\bibitem{Bulck:SEC:18}
J.~Van~Bulck, M.~Minkin, O.~Weisse, D.~Genkin, B.~Kasikci, F.~Piessens,
  M.~Silberstein, T.~F. Wenisch, Y.~Yarom, and R.~Strackx, ``Foreshadow:
  Extracting the keys to the {I}ntel {SGX} kingdom with transient out-of-order
  execution,'' in \emph{Proceedings of the 27th {USENIX} Security Symposium},
  Aug. 2018.

\bibitem{Chen:EuroSP:19}
G.~Chen, S.~Chen, Y.~Xiao, Y.~Zhang, Z.~Lin, and T.~H. Lai, ``{SgxPectre}:
  Stealing {Intel} secrets from {SGX} enclaves via speculative execution,'' in
  \emph{Proceedings of the 4th {IEEE} European Symposium on Security and
  Privacy ({EuroS\&P})}, Jun. 2019.

\bibitem{Murdock:SP:20}
K.~Murdock, D.~Oswald, F.~D. Garcia, J.~Van~Bulck, D.~Gruss, and F.~Piessens,
  ``Plundervolt: Software-based fault injection attacks against {Intel}
  {SGX},'' in \emph{Proceedings of the 2020 {IEEE} Symposium on Security and
  Privacy ({S\&P})}, May 2020.

\bibitem{Alder:ACSAC:20}
F.~Alder, J.~Van~Bulck, D.~Oswald, and F.~Piessens, ``Faulty point unit: {ABI}
  poisoning attacks on {Intel} {SGX},'' in \emph{Proceedings of the 36th Annual
  Computer Security Applications Conference ({ACSAC})}, Dec. 2020, pp.
  415--427.

\bibitem{Ahmad:NDSS:18}
A.~Ahmad, K.~Kim, M.~I. Sarfaraz, and B.~Lee, ``{OBLIVIATE}: A data oblivious
  filesystem for {Intel} {SGX},'' in \emph{Proceedings of the 25th Annual
  Network and Distributed System Security Symposium ({NDSS})}.\hskip 1em plus
  0.5em minus 0.4em\relax ISOC, Feb. 2018.

\bibitem{Oleksenko:ATC:18}
O.~Oleksenko, B.~Trach, R.~Krahn, A.~Martin, C.~Fetzer, and M.~Silberstein,
  ``Varys: Protecting {SGX} enclaves from practical side-channel attacks,'' in
  \emph{Proceedings of the 2018 {USENIX} Annual Technical Conference ({ATC})},
  Jul. 2018, pp. 227--239.

\bibitem{Brasser:ACSAC:19}
F.~Brasser, S.~Capkun, A.~Dmitrienko, T.~Frassetto, K.~Kostiainen, and A.-R.
  Sadeghi, ``{DR.SGX}: Automated and adjustable side-channel protection for
  {SGX} using data location randomization,'' in \emph{Proceedings of the 35th
  Annual Computer Security Applications Conference ({ACSAC})}, Dec. 2019, pp.
  788--800.

\bibitem{Jauernig:M:SP:20}
P.~Jauernig, A.-R. Sadeghi, and E.~Stapf, ``Trusted execution environments:
  Properties, applications, and challenges,'' \emph{{IEEE} Security \&
  Privacy}, vol.~18, no.~2, pp. 56--60, Mar. 2020.

\bibitem{Martin:TR:08:IntroTC}
A.~Martin, ``The ten-page introduction to trusted computing,'' University of
  Oxford, Tech. Rep., Nov. 2008.

\bibitem{Beekman:AsiaCCS:16}
J.~G. Beekman, J.~L. Manferdelli, and D.~Wagner, ``Attestation transparency:
  Building secure internet services for legacy clients,'' in \emph{Proceedings
  of the 11th {ACM} Asia Conference on Computer and Communications Security
  ({Asia CCS})}, May 2016, pp. 687--698.

\bibitem{Anati:HASP:13}
I.~Anati, S.~Gueron, S.~Johnson, and V.~Scarlata, ``Innovative technology for
  {CPU} based attestation and sealing,'' in \emph{Proceedings of the 2nd
  International Workshop on Hardware and Architectural Support for Security and
  Privacy ({HASP})}, Jun. 2013.

\bibitem{Intel:20:IASr6}
\emph{Attestation Service for Intel Software Guard Extensions: {API}
  Documentation}, Intel Corporation, 2020, revision 6.0.

\bibitem{Lowe:CSFW:97}
G.~Lowe, ``A hierarchy of authentication specifications,'' in \emph{Proceedings
  of the 10th Computer Security Foundations Workshop ({CSFW})}, Jun. 1997, pp.
  31--43.

\bibitem{Intel:16:EPID}
S.~Johnson, V.~Scarlata, C.~Rozas, E.~Brickell, and F.~McKeen, \emph{{Intel}
  Software Guard Extensions: {EPID} Provisioning and Attestation Services},
  Intel Corporation, 2016.

\bibitem{Intel:18:DCAP}
V.~Scarlata, S.~Johnson, J.~Beaney, and P.~Zmijewski, \emph{Supporting Third
  Party Attestation for {Intel} {SGX} with {I}ntel Data Center Attestation
  Primitives}, Intel Corporation, 2018.

\bibitem{Chen:CCS:19}
G.~Chen, Y.~Zhang, and T.-H. Lai, ``{OPERA}: Open remote attestation for
  {I}ntel's secure enclaves,'' in \emph{Proceedings of the 2019 {ACM} {SIGSAC}
  Conference on Computer and Communications Security ({CCS})}, Nov. 2019.

\bibitem{Ambrosin:COMST:20}
M.~Ambrosin, M.~Conti, R.~Lazzeretti, M.~M. Rabbani, and S.~Ranise,
  ``Collective remote attestation at the internet of things scale:
  State-of-the-art and future challenges,'' \emph{{IEEE} Communications Surveys
  \& Tutorials}, vol.~22, no.~4, pp. 2447--2461, Jul. 2020.

\bibitem{Chen:SEC:22}
G.~Chen and Y.~Zhang, ``{MAGE}: Mutual attestation for a group of enclaves
  without trusted third parties,'' in \emph{Proceedings of the 31st {USENIX}
  Security Symposium}, Aug. 2022.

\bibitem{Andrade:PRDC:23}
D.~Andrade, J.~Silva, and M.~Correia, ``I can't escape myself: Cloud
  inter-processor attestation and sealing using {Intel} {SGX},'' in
  \emph{Proceedings of the 28th {IEEE} Pacific Rim International Symposium on
  Dependable Computing ({PRDC})}, Oct. 2023, pp. 198--208.

\bibitem{Teaclave:22:web:MA}
{Apache Software Foundation}, ``Mutual attestation: Why and how -- {Apache}
  {Teaclave} (incubating),''
  \url{https://teaclave.apache.org/docs/mutual-attestation/}, Jun. 2023,
  online, Accessed: 2024-09-02.

\bibitem{rfc9334}
H.~Birkholz, D.~Thaler, M.~Richardson, N.~Smith, and W.~Pan, ``Remote
  attestation procedures ({RATS}) architecture,''
  \url{http://www.rfc-editor.org/rfc/rfc9334.txt}, {RFC} Editor, Informational
  {RFC} 9334, Jan. 2023.

\bibitem{Bishop:CompSystems:96}
M.~Bishop and M.~Dilger, ``Checking for race conditions in file accesses,''
  \emph{Computing Systems}, vol.~9, no.~2, pp. 131--152, 1996.

\bibitem{Nunes:CCS:21}
I.~de~Oliveira~Nunes, S.~Jakkamsetti, N.~Rattanavipanon, and G.~Tsudik, ``On
  the {TOCTOU} problem in remote attestation,'' in \emph{Proceedings of the
  2021 {ACM} {SIGSAC} Conference on Computer and Communications Security
  ({CCS})}, Nov. 2021.

\bibitem{Ménétrey:ICDCS:22}
J.~Ménétrey, M.~Pasin, P.~Felber, and V.~Schiavoni, ``{WaTZ:} a trusted
  {WebAssembly} runtime environment with remote attestation for {TrustZone},''
  in \emph{Proceedings of the 42nd {IEEE} International Conference on
  Distributed Computing Systems ({ICDCS})}, Jul. 2022.

\end{thebibliography}

\appendices
\section{Topological Patterns}
\label{sec:topologies}

This section discusses existing topologies for RA, adopts these topologies for MA, and proposes a new topology called delegation model. The topology adopted for the \hema protocol (\sref{sec:hema:protocol}) is in \autoref{fig:hema:tp-ma-111} and \autoref{fig:hema:tp-ma-112}.

\subsection{Remote Attestation}
\label{sec:hera:tp}

The three main roles -- Attester (A), Relying Party (P), and Verifier (V) -- can be classified in different topologies according to their data flows. This section presents three examples that can be used as reference models. These are the Passport Model and the Background-Check Model, which are originally described in RFC 9334~\cite{rfc9334}, and the Delegation Model which is new. For each model there is an abstract version, which doesn't take freshness into account, showing the exchange of the Attestation Evidence (E) and the Attestation Report (R); and a version that includes an Attestation Challenge (C) for freshness. The use of nonces for freshness, instead of, for example, time-based approaches, is due to its simplicity.

\autoref{fig:hema:modelos-ra} shows the topological patterns for remote attestation. The subfigures (\subref{fig:hema:modelo-ra-10}, \subref{fig:hema:modelo-ra-20}, \subref{fig:hema:modelo-ra-30}), on the left side, are the abstract versions with no freshness, and the subfigures (\subref{fig:hema:modelo-ra-11}, \subref{fig:hema:modelo-ra-21}, \subref{fig:hema:modelo-ra-31}), on the right side, depict updated versions with a nonce-based approach.

\begin{figure}
	\centering
	\begin{subfigure}[t]{0.465\linewidth}
		\centering
		\begin{tikzpicture}[
	node distance=1cm,
	-{Latex[scale width=1.5,harpoon]},
	role/.style={
		draw, circle,
		minimum size=1cm, text height=1.5ex, text depth=0.25ex,
		top color=white, bottom color=black!10,
		thick,
	}
]
	\node [role] (A) {A};
	\node [role] (P) [right=of A] {P};
	\node [role] (V) [above=of A] {V};

	\draw ($(A.north)+(-0.1,0)$) -- node [at start] {\numd{1}} node [anchor=east] {E} ($(V.south)+(-0.1,0)$);
	\draw ($(V.south)+(+0.1,0)$) -- node [at start] {\numd{2}} node [anchor=west] {R} ($(A.north)+(+0.1,0)$);
	\draw (A) -- node [at start] {\numd{3}} node [anchor=south] {R} (P);

	\path node [fit=(A)(V)(P), inner sep=1ex] {};
\end{tikzpicture}
		\caption{RA Passport Model (RFC 9334)}
		\label{fig:hema:modelo-ra-10}
	\end{subfigure}
	\quad
	\begin{subfigure}[t]{0.465\linewidth}
		\centering
		\begin{tikzpicture}[
	node distance=1cm,
	-{Latex[scale width=1.5,harpoon]},
	role/.style={
		draw, circle,
		minimum size=1cm, text height=1.5ex, text depth=0.25ex,
		top color=white, bottom color=black!10,
		thick,
	}
]
	\node [role] (A) {A};
	\node [role] (P) [right=of A] {P};
	\node [role] (V) [above=of A] {V};

	\draw ($(A.north)+(-0.1,0)$) -- node [at start] {\numd{2}} node [anchor=east] {E} ($(V.south)+(-0.1,0)$);
	\draw ($(V.south)+(+0.1,0)$) -- node [at start] {\numd{3}} node [anchor=west] {R} ($(A.north)+(+0.1,0)$);

	\draw [-{Latex[scale width=1.5,harpoon,swap]}] ($(P.west)+(0,0.1)$) -- node [at start] {\numd{1}} node [anchor=south] {C} ($(A.east)+(0,0.1)$);
	\draw [-{Latex[scale width=1.5,harpoon,swap]}] ($(A.east)+(0,-0.1)$) -- node [at start] {\numd{4}} node [anchor=north] {R} ($(P.west)+(0,-0.1)$);

	\path node [fit=(A)(V)(P), inner sep=1ex] {};
\end{tikzpicture}
		\caption{RA Passport Model with Attestation Challenge}
		\label{fig:hema:modelo-ra-11}
	\end{subfigure}
	\par\medskip
	\begin{subfigure}[t]{0.465\linewidth}
		\centering
		\begin{tikzpicture}[
	node distance=1cm,
	-{Latex[scale width=1.5,harpoon]},
	role/.style={
		draw, circle,
		minimum size=1cm, text height=1.5ex, text depth=0.25ex,
		top color=white, bottom color=black!10,
		thick,
	}
]
	\node [role] (A) {A};
	\node [role] (P) [right=of A] {P};
	\node [role] (V) [above=of P] {V};

	\draw (A.east) -- node [at start] {\numd{1}} node [anchor=north] {E} (P.west);

	\draw ($(P.north)+(-0.1,0)$) -- node [at start] {\numd{2}} node [anchor=east] {E} ($(V.south)+(-0.1,0)$);
	\draw ($(V.south)+(0.1,0)$) -- node [at start] {\numd{3}} node [anchor=west] {R} ($(P.north)+(0.1,0)$);

	\path node [fit=(A)(V)(P), inner sep=1ex] {};
\end{tikzpicture}
		\caption{RA Background-Check Model (RFC 9334)}
		\label{fig:hema:modelo-ra-20}
	\end{subfigure}
	\quad
	\begin{subfigure}[t]{0.465\linewidth}
		\centering
		\begin{tikzpicture}[
	node distance=1cm,
	-{Latex[scale width=1.5,harpoon]},
	role/.style={
		draw, circle,
		minimum size=1cm, text height=1.5ex, text depth=0.25ex,
		top color=white, bottom color=black!10,
		thick,
	}
]
	\node [role] (A) {A};
	\node [role] (P) [right=of A] {P};
	\node [role] (V) [above=of P] {V};

	\draw [-{Latex[scale width=1.5,harpoon,swap]}] ($(P.west)+(0,0.1)$) -- node [at start] {\numd{1}} node [anchor=south] {C} ($(A.east)+(0,0.1)$);
	\draw [-{Latex[scale width=1.5,harpoon,swap]}] ($(A.east)+(0,-0.1)$) -- node [at start] {\numd{2}} node [anchor=north] {E} ($(P.west)+(0,-0.1)$);

	\draw ($(P.north)+(-0.1,0)$) -- node [at start] {\numd{3}} node [anchor=east] {E} ($(V.south)+(-0.1,0)$);
	\draw ($(V.south)+(0.1,0)$) -- node [at start] {\numd{4}} node [anchor=west] {R} ($(P.north)+(0.1,0)$);

	\path node [fit=(A)(V)(P), inner sep=1ex] {};
\end{tikzpicture}
		\caption{RA Background-Check Model with Attestation Challenge}
		\label{fig:hema:modelo-ra-21}
	\end{subfigure}
	\par\medskip
	\begin{subfigure}[t]{0.465\linewidth}
		\centering
		\begin{tikzpicture}[
	node distance=1cm,
	-{Latex[scale width=1.5,harpoon]},
	role/.style={
		draw, circle,
		minimum size=1cm, text height=1.5ex, text depth=0.25ex,
		top color=white, bottom color=black!10,
		thick,
	}
]
	\node [role] (A) {A};
	\node [role] (V) [right=of A] {V};
	\node [role] (P) [right=of V] {P};

	\draw (A.east) -- node [at start] {\numd{1}} node [anchor=south] {E} (V.west);
	\draw (V.east) -- node [at start] {\numd{2}} node [anchor=south] {R} (P.west);

	\path node [fit=(A)(V)(P), inner sep=1ex] {};
\end{tikzpicture}
		\caption{RA Delegation Model}
		\label{fig:hema:modelo-ra-30}
	\end{subfigure}
	\quad
	\begin{subfigure}[t]{0.465\linewidth}
		\centering
		\begin{tikzpicture}[
	node distance=1cm,
	-{Latex[scale width=1.5,harpoon]},
	role/.style={
		draw, circle,
		minimum size=1cm, text height=1.5ex, text depth=0.25ex,
		top color=white, bottom color=black!10,
		thick,
	}
]
	\node [role] (A) {A};
	\node [role] (V) [right=of A] {V};
	\node [role] (P) [right=of V] {P};

	\draw ($(P.west)+(0,-0.1)$) -- node [at start] {\numd{1}} node [anchor=north] {C} ($(V.east)+(0,-0.1)$);
	\draw ($(V.west)+(0,-0.1)$) -- node [at start] {\numd{2}} node [anchor=north] {C} ($(A.east)+(0,-0.1)$);
	\draw ($(A.east)+(0,+0.1)$) -- node [at start] {\numd{3}} node [anchor=south] {E} ($(V.west)+(0,+0.1)$);
	\draw ($(V.east)+(0,+0.1)$) -- node [at start] {\numd{4}} node [anchor=south] {R} ($(P.west)+(0,+0.1)$);

	\path node [fit=(A)(V)(P), inner sep=1ex] {};
\end{tikzpicture}
		\caption{RA Delegation Model with Attestation Challenge}
		\label{fig:hema:modelo-ra-31}
	\end{subfigure}
	\caption[Topological patterns for RA]{Three topological patterns for RA. The two reference models (\subref{fig:hema:modelo-ra-10}, \subref{fig:hema:modelo-ra-20}) from RFC 9334, our RA reference model (\subref{fig:hema:modelo-ra-30}), and the three updated versions that include an Attestation Challenge (\subref{fig:hema:modelo-ra-11}, \subref{fig:hema:modelo-ra-21}, \subref{fig:hema:modelo-ra-31}).}
	\label{fig:hema:modelos-ra}
\end{figure}

In the passport model, the Attester sits between the Relying Party and the Verifier and handles all interaction with the Verifier.
In \autoref{fig:hema:modelo-ra-10}, the Attester sends the Attestation Evidence to the Verifier and receives, in return, the Attestation Report which it forwards to the Relying Party.
In \autoref{fig:hema:modelo-ra-11}, the Attester receives an Attestation Challenge from the Relying Party, sends the Attestation Evidence to the Verifier, and receives an Attestation Report, in return, which it forwards to the Relying Party for appraisal.

The simpler version of the passport model allows the Attester to reuse the Attestation Report, delivering it to different Relying Parties. With nonces, however, this does not work because each Relying Party expects to find its own nonce in the Attestation Report.


In the background-check model, the Relying Party sits between the Attester and the Verifier and handles all the interaction with the Verifier.
In \autoref{fig:hema:modelo-ra-20}, the Attester sends the Attestation Evidence to the Relying Party, which forwards it to the Verifier and receives, in return, the Attestation Report.
In \autoref{fig:hema:modelo-ra-21}, the Attester receives an Attestation Challenge from the Relying Party and replies with the Attestation Evidence. The Relying Party sends the Attestation Evidence to the Verifier and receives the Attestation Report in return.

The background-check model is a good option when the Attester has no access to the Verifier. This could happen, for example, on a network where only the Relying Party is granted access to external services, or when the Attester does not have the software stack to communicate with the Verifier.


In the delegation model, the Relying Party delegates the interaction with the Attester to the Verifier, and receives an Attestation Report at the end of the data flow.
In \autoref{fig:hema:modelo-ra-31}, the Attester sends the Attestation Evidence to the Verifier. The Verifier appraises the Attestation Evidence and sends the Attestation Report to the Relying Party.
In \autoref{fig:hema:modelo-ra-31}, the Verifier receives an Attestation Challenge from the Relying Party, and forwards the Attestation Challenge to the Attester. The Attester replies to the Verifier with the Attestation Evidence and the Verifier sends to the Relying Party the Attestation Report for appraisal.

The delegation model could have the Attester configured in a such a way as to send continuous attestation updates to the Relying Party. Every amount of time the Attester would send the Attestation Evidence to the Verifier, which appraises the evidence and sends the Attestation Report to the Relying Party. The Relying Party expects to receive an Attestation Report for a particular Attester regularly. If the Relying Party doesn't receive the Attestation Report within the expected period it can decide to take action such as mark this Attester as untrustworthy and/or unavailable.


The Attester needs a way to trigger the attestation protocol. The most obvious would be having a source of time, which in this case does not need to be trusted. The trustworthiness of the Attester is not compromised even if the adversary compromises the clock. The adversary can increase the clock, in which case the attestation protocol is triggered more frequently, or can decrease the clock, in which case the attestation protocol is triggered less frequently. If the former happens it increases resource usage but it does not compromise the Attester, and if the latter happens the Relying Party eventually notices the Attester missed its regular attestation and can take action in order to mitigate possible adverse effects to the system.

The attestation frequency depends on policy. The higher the attestation frequency the higher the resource consumption. On the other hand, having a low attestation frequency increases the risk of compromise since the results in the Attestation Report are valid only for the specific point in time when the attestation occurred. The Relying Party cannot infer, from a set of Attestation Reports, whether the Attester is healthy or compromised during the intervals between attestations. This problem is known as time-of-check to time-of-use (TOCTOU)~\cite{Bishop:CompSystems:96, Nunes:CCS:21}. In general, a TOCTOU occurs when a resource is checked at time $t_{1}$ and an action is taken at time $t_{2}$ assuming the previous check, carried during $t_{1}$, is still valid during $t_{2}$ when this is not the case. Increasing the attestation frequency reduces the vulnerability window.

While it is common to have the Attester, the Relying Party, and the Verifier separate this is not a fixed rule~\cite{Ménétrey:ICDCS:22}. The three main roles and their organization, and how freshness is dealt with, if at all, comes down to architectural choices.

\subsubsection{Cost}

\autoref{hera:tab:modelos} compares the cost of the different topological patterns in regard to the total number of messages exchanged and the amount of channels required. The communication channels for these topological patterns are half-duplex, that is, the channels are bidirectional but communication happens one direction at a time.

The Passport Model and the Background-Check Model have the same cost of three messages exchanged with the Delegation Model being the less costly reference model with two messages exchanged. The three versions that include a challenge have the same cost of four messages exchanged. The number of communication channels required is two for all reference models.

\begin{table}
	\caption{Cost of the topological patterns in \hera.}
	\label{hera:tab:modelos}
	\centering
	\begin{tabulary}{\textwidth}{@{}Lccc@{}}
		\toprule
		Model & \#Messages & \#Channels & Trigger \\
		\midrule
		\autoref{fig:hema:modelo-ra-10} & 3 & 2 & A \\
		\autoref{fig:hema:modelo-ra-20} & 3 & 2 & A \\
		\autoref{fig:hema:modelo-ra-30} & 2 & 2 & A \\
		\autoref{fig:hema:modelo-ra-11} & 4 & 2 & P \\
		\autoref{fig:hema:modelo-ra-21} & 4 & 2 & P \\
		\autoref{fig:hema:modelo-ra-31} & 4 & 2 & P \\
		\bottomrule
	\end{tabulary}
\end{table}

\subsubsection{Freshness}

Freshness can be considered from three different perspectives. The Verifier may want to know whether the Attestation Evidence is fresh. The Attester may want to know whether the Attestation Report is fresh. The Relying Party may want to know whether both the Attestation Evidence and the Attestation Report are fresh. However, in our reference models the Relying Party is the only role issuing a challenge. The reason behind this is that it is the Relying Party that consumes and appraises the Attestation Report before deciding whether to service its clients, that is, the Attesters.

The Attester considers the Attestation Report as an opaque blob and is therefore not concerned with its freshness. The Verifier examines a set of claims contained within the Attestation Evidence and produces an Attestation Report based on those claims, but without itself using the results of the attestation. Whether the claims in the Attestation Evidence are fresh is inconsequential to the Verifier, in our reference models, and it is the party using the Attestation Report that must ensure this is the case.

The Attestation Challenge is part of the Attestation Evidence which in turn is part of the Attestation Report. The Relying Party ensures the freshness of both the Attestation Evidence and the Attestation Report by having the Attester include the Relying Party's Attestation Challenge in the Attestation Evidence.

\autoref{fig:hera:sd:0} and \autoref{fig:hera:sd:1} rely on nonces for freshness but other alternatives could be used such as time-based approaches~\cite[Appendix~A]{rfc9334}. Time-based approaches, however, require a trusted time source and a clock synchronization mechanism which may not be readily available within a TEE.

\subsection{Mutual Attestation}
\label{sec:hema:tp}

\autoref{fig:hema:modelos-ma} shows topological patterns for the MA protocol that can be used as reference models. We used circles for RA, in the previous section, and use square for MA, in this section, to avoid confusion. The topological patterns for RA, in \aref{sec:hera:tp}, all have one Verifier because there is a single Attester whereas the topological patterns for MA, in this section, may have one Verifier or two Verifiers. The data flow involves two Attesters and either one Verifier or two Verifiers and focus on the exchange of Attestation Challenges, Attestation Evidence, and Attestation Reports.
The Attesters send Attestation Evidence, which contains the Attestation Challenges of the Initiator and the Responder, to the Verifiers for appraisal and receive, in return, Attestation Reports which they exchange with one another. This order of actions constrains the exchange of messages since the Attestation Report requires having the Attestation Evidence, and the Attestation Evidence requires having both nonces. The Initiator always triggers the communication, which means the Responder does not start its part of the protocol before receiving a first message. The number of messages exchanged between the different parties is minimized.

\begin{figure*}[htbp]
	\centering
	\begin{subfigure}[t]{0.22\linewidth}
		\centering
		\begin{tikzpicture}[
	node distance=1cm,
	-{Latex[scale width=1.5,harpoon]},
	role1/.style={
		draw, circle,
		minimum size=1cm, text height=1.5ex, text depth=0.25ex,
		top color=white, bottom color=black!10,
		thick,
	},
	role2/.style={
		draw, rectangle,
		minimum size=1cm,
		minimum height=2cm,
		text height=1.5ex, text depth=0.25ex,
		top color=white, bottom color=black!10,
		thick,
	},
	role3/.style={
		draw, rectangle,
		minimum size=2cm,
		text height=1.5ex, text depth=0.25ex,
		top color=white, bottom color=black!10,
		thick,
	},
	every fit/.style={inner sep=3ex},
]
	\node [role3] (V) at (2,3) {V};
	\node [role3] (I) at (0,0) {I};
	\node [role3] (R) at (4,0) {R};

	\draw ($(I.east)+(0,0.15)$) -- node [at start] {\numd{3}} node [anchor=south] {$R_{1}$} ($(R.west)+(0,0.15)$);
	\draw ($(R.west)+(0,-0.15)$) -- node [at start] {\numd{6}} node [anchor=north] {$R_{2}$} ($(I.east)+(0,-0.15)$);

	\draw ($(I.north)+(-0.15,0)$) |- node [at start] {\numd{1}} node [anchor=east] {$E_{1}$} ($(V.west)+(0,+0.15)$);
	\draw ($(V.west)+(0,-0.15)$) -| node [at start] {\numd{2}} node [anchor=north west] {$R_{1}$} ($(I.north)+(0.15,0)$);

	\draw ($(R.north)+(-0.15,0)$) |- node [at start] {\numd{4}} node [anchor=north east] {$E_{2}$} ($(V.east)+(0,-0.15)$);
	\draw ($(V.east)+(0,0.15)$) -| node [at start] {\numd{5}} node [anchor=west] {$R_{2}$} ($(R.north)+(0.15,0)$);

	\node [fit=(V)(I)(R)] {}; 
\end{tikzpicture}
		\caption{MA Passport Model}
		\label{fig:hema:tp-ma-101}
	\end{subfigure}
	\hfill
	\begin{subfigure}[t]{0.22\linewidth}
		\centering
		\begin{tikzpicture}[
	node distance=1cm,
	-{Latex[scale width=1.5,harpoon]},
	role1/.style={
		draw, circle,
		minimum size=1cm, text height=1.5ex, text depth=0.25ex,
		top color=white, bottom color=black!10,
		thick,
	},
	role2/.style={
		draw, rectangle,
		minimum size=1cm,
		minimum height=2cm,
		text height=1.5ex, text depth=0.25ex,
		top color=white, bottom color=black!10,
		thick,
	},
	role3/.style={
		draw, rectangle,
		minimum size=2cm,
		text height=1.5ex, text depth=0.25ex,
		top color=white, bottom color=black!10,
		thick,
	},
	every fit/.style={inner sep=3ex},
]
	\node [role3] (V) at (2,3) {V};
	\node [role3] (I) at (0,0) {I};
	\node [role3] (R) at (4,0) {R};

	\draw ($(I.east)+(0,0.8)$) -- node [at start] {\numd{1}} node [anchor=south] {$C_{1}$} ($(R.west)+(0,0.8)$);
	\draw ($(R.west)+(0,0.5)$) -- node [at start] {\numd{4}} node [anchor=north] {$C_{2},R_{2}$} ($(I.east)+(0,0.5)$);
	\draw ($(I.east)+(0,-0.5)$) -- node [at start] {\numd{7}} node [anchor=south] {$R_{1}$} ($(R.west)+(0,-0.5)$);

	\draw ($(I.north)+(-0.15,0)$) |- node [at start] {\numd{5}} node [anchor=east] {$E_{1}$} ($(V.west)+(0,+0.15)$);
	\draw ($(V.west)+(0,-0.15)$) -| node [at start] {\numd{6}} node [anchor=north west] {$R_{1}$} ($(I.north)+(0.15,0)$);

	\draw ($(R.north)+(-0.15,0)$) |- node [at start] {\numd{2}} node [anchor=north east] {$E_{2}$} ($(V.east)+(0,-0.15)$);
	\draw ($(V.east)+(0,0.15)$) -| node [at start] {\numd{3}} node [anchor=west] {$R_{2}$} ($(R.north)+(0.15,0)$);

	\begin{scope}[on background layer]
		\node [fill=orange!15, fit=(V)(I)(R)] {};
	\end{scope}
\end{tikzpicture}
		\caption{MA Passport Model with Attestation Challenge}
		\label{fig:hema:tp-ma-111}
	\end{subfigure}
	\hfill
	\begin{subfigure}[t]{0.22\linewidth}
		\centering
		\begin{tikzpicture}[
	node distance=1cm,
	-{Latex[scale width=1.5,harpoon]},
	role1/.style={
		draw, circle,
		minimum size=1cm, text height=1.5ex, text depth=0.25ex,
		top color=white, bottom color=black!10,
		thick,
	},
	role2/.style={
		draw, rectangle,
		minimum size=1cm,
		minimum height=2cm,
		text height=1.5ex, text depth=0.25ex,
		top color=white, bottom color=black!10,
		thick,
	},
	role3/.style={
		draw, rectangle,
		minimum size=2cm,
		text height=1.5ex, text depth=0.25ex,
		top color=white, bottom color=black!10,
		thick,
	},
	every fit/.style={inner sep=3ex},
]
	\node [role3] (U) at (0,4) {V1};
	\node [role3] (V) at (4,4) {V2};
	\node [role3] (I) at (0,0) {I};
	\node [role3] (R) at (4,0) {R};

	\draw ($(I.north)+(-0.15,0)$) -- node [at start] {\numd{1}} node [anchor=east] {$E_{1}$} ($(U.south)+(-0.15,0)$);
	\draw ($(U.south)+(+0.15,0)$) -- node [at start] {\numd{2}} node [anchor=west] {$R_{1}$} ($(I.north)+(0.15,0)$);

	\draw ($(I.east)+(0,0.15)$) -- node [at start] {\numd{3}} node [anchor=south] {$R_{1}$} ($(R.west)+(0,0.15)$);
	\draw ($(R.west)+(0,-0.15)$) -- node [at start] {\numd{6}} node [anchor=north] {$R_{2}$} ($(I.east)+(0,-0.15)$);

	\draw ($(R.north)+(-0.15,0)$) -- node [at start] {\numd{4}} node [anchor=east] {$E_{2}$} ($(V.south)+(-0.15,0)$);
	\draw ($(V.south)+(+0.15,0)$) -- node [at start] {\numd{5}} node [anchor=west] {$R_{2}$} ($(R.north)+(0.15,0)$);

	\node [fit=(U)(V)(I)(R)] {}; 
\end{tikzpicture}
		\caption{MA Passport Model}
		\label{fig:hema:tp-ma-102}
	\end{subfigure}
	\hfill
	\begin{subfigure}[t]{0.22\linewidth}
		\centering
		\begin{tikzpicture}[
	node distance=1cm,
	-{Latex[scale width=1.5,harpoon]},
	role1/.style={
		draw, rectangle,
		minimum size=1cm, text height=1.5ex, text depth=0.25ex,
		top color=white, bottom color=black!10,
		thick,
	},
	role2/.style={
		draw, rectangle,
		minimum size=1cm,
		minimum height=2cm,
		text height=1.5ex, text depth=0.25ex,
		top color=white, bottom color=black!10,
		thick,
	},
	role3/.style={
		draw, rectangle,
		minimum size=2cm,
		text height=1.5ex, text depth=0.25ex,
		top color=white, bottom color=black!10,
		thick,
	},
	every fit/.style={inner sep=3ex},
]
	\node [role3] (U) at (0,4) {V1};
	\node [role3] (V) at (4,4) {V2};
	\node [role3] (I) at (0,0) {I};
	\node [role3] (R) at (4,0) {R};

	\draw ($(I.east)+(0,0.8)$) -- node [at start] {\numd{1}} node [anchor=south] {$C_{1}$} ($(R.west)+(0,0.8)$);
	\draw ($(R.west)+(0,0.5)$) -- node [at start] {\numd{4}} node [anchor=north] {$C_{2},R_{2}$} ($(I.east)+(0,0.5)$);
	\draw ($(I.east)+(0,-0.5)$) -- node [at start] {\numd{7}} node [anchor=south] {$R_{1}$} ($(R.west)+(0,-0.5)$);

	\draw ($(I.north)+(-0.15,0)$) -- node [at start] {\numd{5}} node [anchor=east] {$E_{1}$} ($(U.south)+(-0.15,0)$);
	\draw ($(U.south)+(+0.15,0)$) -- node [at start] {\numd{6}} node [anchor=west] {$R_{1}$} ($(I.north)+(0.15,0)$);

	\draw ($(R.north)+(-0.15,0)$) -- node [at start] {\numd{2}} node [anchor=east] {$E_{2}$} ($(V.south)+(-0.15,0)$);
	\draw ($(V.south)+(+0.15,0)$) -- node [at start] {\numd{3}} node [anchor=west] {$R_{2}$} ($(R.north)+(0.15,0)$);

	\begin{scope}[on background layer]
		\node [fill=orange!15, fit=(V)(I)(R)] {};
	\end{scope}
\end{tikzpicture}
		\caption{MA Passport Model with Attestation Challenge}
		\label{fig:hema:tp-ma-112}
	\end{subfigure}
	\par\bigskip
	\begin{subfigure}[t]{0.22\linewidth}
		\centering
		\begin{tikzpicture}[
	node distance=1cm,
	-{Latex[scale width=1.5,harpoon]},
	role1/.style={
		draw, rectangle,
		minimum size=1cm, text height=1.5ex, text depth=0.25ex,
		top color=white, bottom color=black!10,
		thick,
	},
	role2/.style={
		draw, rectangle,
		minimum size=1cm,
		minimum height=2cm,
		text height=1.5ex, text depth=0.25ex,
		top color=white, bottom color=black!10,
		thick,
	},
	role3/.style={
		draw, rectangle,
		minimum size=2cm,
		text height=1.5ex, text depth=0.25ex,
		top color=white, bottom color=black!10,
		thick,
	},
	every fit/.style={inner sep=3ex},
]
	\node [role3] (V) at (2,3) {V};
	\node [role3] (I) at (0,0) {I};
	\node [role3] (R) at (4,0) {R};

	\draw ($(I.east)+(0,0.15)$) -- node [at start] {\numd{1}} node [anchor=south] {$E_{1}$} ($(R.west)+(0,0.15)$);
	\draw ($(R.west)+(0,-0.15)$) -- node [at start] {\numd{4}} node [anchor=north] {$E_{2}$} ($(I.east)+(0,-0.15)$);

	\draw ($(I.north)+(-0.15,0)$) |- node [at start] {\numd{5}} node [anchor=east] {$E_{2}$} ($(V.west)+(0,+0.15)$);
	\draw ($(V.west)+(0,-0.15)$) -| node [at start] {\numd{6}} node [anchor=north west] {$R_{2}$} ($(I.north)+(0.15,0)$);

	\draw ($(R.north)+(-0.15,0)$) |- node [at start] {\numd{2}} node [anchor=north east] {$E_{1}$} ($(V.east)+(0,-0.15)$);
	\draw ($(V.east)+(0,0.15)$) -| node [at start] {\numd{3}} node [anchor=west] {$R_{1}$} ($(R.north)+(0.15,0)$);

	\node [fit=(V)(I)(R)] {}; 
\end{tikzpicture}
		\caption{MA Background-Check Model}
		\label{fig:hema:tp-ma-201}
	\end{subfigure}
	\hfill
	\begin{subfigure}[t]{0.22\linewidth}
		\centering
		\begin{tikzpicture}[
	node distance=1cm,
	-{Latex[scale width=1.5,harpoon]},
	role1/.style={
		draw, rectangle,
		minimum size=1cm, text height=1.5ex, text depth=0.25ex,
		top color=white, bottom color=black!10,
		thick,
	},
	role2/.style={
		draw, rectangle,
		minimum size=1cm,
		minimum height=2cm,
		text height=1.5ex, text depth=0.25ex,
		top color=white, bottom color=black!10,
		thick,
	},
	role3/.style={
		draw, rectangle,
		minimum size=2cm,
		text height=1.5ex, text depth=0.25ex,
		top color=white, bottom color=black!10,
		thick,
	},
	every fit/.style={inner sep=3ex},
]
	\node [role3] (V) at (2,3) {V};
	\node [role3] (I) at (0,0) {I};
	\node [role3] (R) at (4,0) {R};

	\draw ($(I.east)+(0,0.8)$) -- node [at start] {\numd{1}} node [anchor=south] {$C_{1}$} ($(R.west)+(0,0.8)$);
	\draw ($(R.west)+(0,0.5)$) -- node [at start] {\numd{2}} node [anchor=north] {$C_{2},E_{2}$} ($(I.east)+(0,0.5)$);
	\draw ($(I.east)+(0,-0.5)$) -- node [at start] {\numd{5}} node [anchor=south] {$E_{1}$} ($(R.west)+(0,-0.5)$);

	\draw ($(I.north)+(-0.15,0)$) |- node [at start] {\numd{3}} node [anchor=east] {$E_{2}$} ($(V.west)+(0,+0.15)$);
	\draw ($(V.west)+(0,-0.15)$) -| node [at start] {\numd{4}} node [anchor=north west] {$R_{2}$} ($(I.north)+(0.15,0)$);

	\draw ($(R.north)+(-0.15,0)$) |- node [at start] {\numd{6}} node [anchor=north east] {$E_{1}$} ($(V.east)+(0,-0.15)$);
	\draw ($(V.east)+(0,0.15)$) -| node [at start] {\numd{7}} node [anchor=west] {$R_{1}$} ($(R.north)+(0.15,0)$);

	\node [fit=(V)(I)(R)] {}; 
\end{tikzpicture}
		\caption{MA Background-Check Model with Attestation Challenge}
		\label{fig:hema:tp-ma-211}
	\end{subfigure}
	\hfill
	\begin{subfigure}[t]{0.22\linewidth}
		\centering
		\begin{tikzpicture}[
	node distance=1cm,
	-{Latex[scale width=1.5,harpoon]},
	role1/.style={
		draw, circle,
		minimum size=1cm, text height=1.5ex, text depth=0.25ex,
		top color=white, bottom color=black!10,
		thick,
	},
	role2/.style={
		draw, rectangle,
		minimum size=1cm,
		minimum height=2cm,
		text height=1.5ex, text depth=0.25ex,
		top color=white, bottom color=black!10,
		thick,
	},
	role3/.style={
		draw, rectangle,
		minimum size=2cm,
		text height=1.5ex, text depth=0.25ex,
		top color=white, bottom color=black!10,
		thick,
	},
	every fit/.style={inner sep=3ex},
]
	\node [role3] (U) at (0,4) {V1};
	\node [role3] (V) at (4,4) {V2};
	\node [role3] (I) at (0,0) {I};
	\node [role3] (R) at (4,0) {R};

	\draw ($(I.east)+(0,0.15)$) -- node [at start] {\numd{1}} node [anchor=south] {$E_{1}$} ($(R.west)+(0,0.15)$);
	\draw ($(R.west)+(0,-0.15)$) -- node [at start] {\numd{4}} node [anchor=north] {$E_{2}$} ($(I.east)+(0,-0.15)$);

	\draw ($(R.north)+(-0.15,0)$) -- node [at start] {\numd{2}} node [anchor=east] {$E_{1}$} ($(V.south)+(-0.15,0)$);
	\draw ($(V.south)+(+0.15,0)$) -- node [at start] {\numd{3}} node [anchor=west] {$R_{1}$} ($(R.north)+(0.15,0)$);

	\draw ($(I.north)+(-0.15,0)$) -- node [at start] {\numd{5}} node [anchor=east] {$E_{2}$} ($(U.south)+(-0.15,0)$);
	\draw ($(U.south)+(+0.15,0)$) -- node [at start] {\numd{6}} node [anchor=west] {$R_{2}$} ($(I.north)+(0.15,0)$);

	\node [fit=(U)(V)(I)(R)] {}; 
\end{tikzpicture}
		\caption{MA Background-Check Model}
		\label{fig:hema:tp-ma-202}
	\end{subfigure}
	\hfill
	\begin{subfigure}[t]{0.22\linewidth}
		\centering
		\begin{tikzpicture}[
	node distance=1cm,
	-{Latex[scale width=1.5,harpoon]},
	role1/.style={
		draw, rectangle,
		minimum size=1cm, text height=1.5ex, text depth=0.25ex,
		top color=white, bottom color=black!10,
		thick,
	},
	role2/.style={
		draw, rectangle,
		minimum size=1cm,
		minimum height=2cm,
		text height=1.5ex, text depth=0.25ex,
		top color=white, bottom color=black!10,
		thick,
	},
	role3/.style={
		draw, rectangle,
		minimum size=2cm,
		text height=1.5ex, text depth=0.25ex,
		top color=white, bottom color=black!10,
		thick,
	},
	every fit/.style={inner sep=3ex},
]
	\node [role3] (U) at (0,4) {V1};
	\node [role3] (V) at (4,4) {V2};
	\node [role3] (I) at (0,0) {I};
	\node [role3] (R) at (4,0) {R};

	\draw ($(I.east)+(0,0.8)$) -- node [at start] {\numd{1}} node [anchor=south] {$C_{1}$} ($(R.west)+(0,0.8)$);
	\draw ($(R.west)+(0,0.5)$) -- node [at start] {\numd{2}} node [anchor=north] {$C_{2},E_{2}$} ($(I.east)+(0,0.5)$);
	\draw ($(I.east)+(0,-0.5)$) -- node [at start] {\numd{5}} node [anchor=south] {$E_{1}$} ($(R.west)+(0,-0.5)$);

	\draw ($(I.north)+(-0.15,0)$) -- node [at start] {\numd{3}} node [anchor=east] {$E_{2}$} ($(U.south)+(-0.15,0)$);
	\draw ($(U.south)+(+0.15,0)$) -- node [at start] {\numd{4}} node [anchor=west] {$R_{2}$} ($(I.north)+(0.15,0)$);

	\draw ($(R.north)+(-0.15,0)$) -- node [at start] {\numd{6}} node [anchor=east] {$E_{1}$} ($(V.south)+(-0.15,0)$);
	\draw ($(V.south)+(+0.15,0)$) -- node [at start] {\numd{7}} node [anchor=west] {$R_{1}$} ($(R.north)+(0.15,0)$);

	\node [fit=(U)(V)(I)(R)] {}; 
\end{tikzpicture}
		\caption{MA Background-Check Model with Attestation Challenge}
		\label{fig:hema:tp-ma-212}
	\end{subfigure}
	\par\bigskip
	\begin{subfigure}[t]{0.22\linewidth}
		\centering
		\begin{tikzpicture}[
	node distance=1cm,
	-{Latex[scale width=1.5,harpoon]},
	role1/.style={
		draw, rectangle,
		minimum size=1cm, text height=1.5ex, text depth=0.25ex,
		top color=white, bottom color=black!10,
		thick,
	},
	role2/.style={
		draw, rectangle,
		minimum size=1cm,
		minimum height=2cm,
		text height=1.5ex, text depth=0.25ex,
		top color=white, bottom color=black!10,
		thick,
	},
	role3/.style={
		draw, rectangle,
		minimum size=2cm,
		text height=1.5ex, text depth=0.25ex,
		top color=white, bottom color=black!10,
		thick,
	},
	every fit/.style={inner sep=3ex},
]
	\node [role3] (V) at (2,3) {V};
	\node [role3] (I) at (0,0) {I};
	\node [role3] (R) at (4,0) {R};

	\draw ($(I.north)+(-0.15,0)$) |- node [at start] {\numd{1}} node [anchor=east] {$E_{1}$} ($(V.west)+(0,+0.15)$);
	\draw ($(V.west)+(0,-0.15)$) -| node [at start] {\numd{4}} node [anchor=north west] {$R_{2}$} ($(I.north)+(0.15,0)$);

	\draw ($(R.north)+(-0.15,0)$) |- node [at start] {\numd{3}} node [anchor=north east] {$E_{2}$} ($(V.east)+(0,-0.15)$);
	\draw ($(V.east)+(0,0.15)$) -| node [at start] {\numd{2}} node [anchor=west] {$R_{1}$} ($(R.north)+(0.15,0)$);

	\node [fit=(V)(I)(R)] {}; 
\end{tikzpicture}
		\caption{MA Delegation Model}
		\label{fig:hema:tp-ma-401}
	\end{subfigure}
	\hfill
	\begin{subfigure}[t]{0.22\linewidth}
		\centering
		\begin{tikzpicture}[
	node distance=1cm,
	-{Latex[scale width=1.5,harpoon]},
	role1/.style={
		draw, rectangle,
		minimum size=1cm, text height=1.5ex, text depth=0.25ex,
		top color=white, bottom color=black!10,
		thick,
	},
	role2/.style={
		draw, rectangle,
		minimum size=1cm,
		minimum height=2cm,
		text height=1.5ex, text depth=0.25ex,
		top color=white, bottom color=black!10,
		thick,
	},
	role3/.style={
		draw, rectangle,
		minimum size=2cm,
		text height=1.5ex, text depth=0.25ex,
		top color=white, bottom color=black!10,
		thick,
	},
	every fit/.style={inner sep=3ex},
]
	\node [role3] (V) at (2,3) {V};
	\node [role3] (I) at (0,0) {I};
	\node [role3] (R) at (4,0) {R};

	\draw ($(I.north)+(-0.8,0)$) |- node [at start] {\numd{1}} node [anchor=east] {$C_{1}$} ($(V.west)+(0,+0.8)$);
	\draw ($(V.west)+(0,+0.5)$) -| node [at start] {\numd{4}} node [anchor=north west] {$C_{2},R_{2}$} ($(I.north)+(-0.5,0)$);
	\draw ($(I.north)+(+0.5,0)$) |- node [at start] {\numd{5}} node [anchor=east] {$E_{1}$} ($(V.west)+(0,-0.5)$);

	\draw ($(V.east)+(0,+0.8)$) -| node [at start] {\numd{2}} node [anchor=west] {$C_{1}$} ($(R.north)+(+0.8,0)$);
	\draw ($(R.north)+(+0.5,0)$) |- node [at start] {\numd{3}} node [anchor=north east] {$C_{2},E_{2}$} ($(V.east)+(0,+0.5)$);
	\draw ($(V.east)+(0,-0.5)$) -| node [at start] {\numd{6}} node [anchor=west] {$R_{1}$} ($(R.north)+(-0.5,0)$);

	\node [fit=(V)(I)(R)] {}; 
\end{tikzpicture}
		\caption{MA Delegation Model with Attestation Challenge}
		\label{fig:hema:tp-ma-411}
	\end{subfigure}
	\hfill
	\begin{subfigure}[t]{0.22\linewidth}
		\centering
		\begin{tikzpicture}[
	node distance=1cm,
	-{Latex[scale width=1.5,harpoon]},
	role1/.style={
		draw, rectangle,
		minimum size=1cm, text height=1.5ex, text depth=0.25ex,
		top color=white, bottom color=black!10,
		thick,
	},
	role2/.style={
		draw, rectangle,
		minimum size=1cm,
		minimum height=2cm,
		text height=1.5ex, text depth=0.25ex,
		top color=white, bottom color=black!10,
		thick,
	},
	role3/.style={
		draw, rectangle,
		minimum size=2cm,
		text height=1.5ex, text depth=0.25ex,
		top color=white, bottom color=black!10,
		thick,
	},
	every fit/.style={inner sep=3ex},
]
	\node [role3] (U) at (2,3) {V1};
	\node [role3] (V) at (2,-3) {V2};
	\node [role3] (I) at (0,0) {I};
	\node [role3] (R) at (4,0) {R};

	\draw (I.north) |- node [at start] {\numd{1}} node [anchor=south] {$E_{1}$} (U.west);
	\draw (U.east) -| node [at start] {\numd{2}} node [anchor=south] {$R_{1}$} (R.north);
	\draw (R.south) |- node [at start] {\numd{3}} node [anchor=north] {$E_{2}$} (V.east);
	\draw (V.west) -| node [at start] {\numd{4}} node [anchor=north] {$R_{2}$} (I.south);

	\node [fit=(U)(V)(I)(R)] {}; 
\end{tikzpicture}
		\caption{MA Delegation Model}
		\label{fig:hema:tp-ma-402}
	\end{subfigure}
	\hfill
	\begin{subfigure}[t]{0.22\linewidth}
		\centering
		\begin{tikzpicture}[
	node distance=1cm,
	-{Latex[scale width=1.5,harpoon]},
	role1/.style={
		draw, rectangle,
		minimum size=1cm, text height=1.5ex, text depth=0.25ex,
		top color=white, bottom color=black!10,
		thick,
	},
	role2/.style={
		draw, rectangle,
		minimum size=1cm,
		minimum height=2cm,
		text height=1.5ex, text depth=0.25ex,
		top color=white, bottom color=black!10,
		thick,
	},
	role3/.style={
		draw, rectangle,
		minimum size=2cm,
		text height=1.5ex, text depth=0.25ex,
		top color=white, bottom color=black!10,
		thick,
	},
	every fit/.style={inner sep=3ex},
]
	\node [role3] (U) at (2,3) {V1};
	\node [role3] (V) at (2,-3) {V2};
	\node [role3] (I) at (0,0) {I};
	\node [role3] (R) at (4,0) {R};

	\draw ($(I.north)+(-0.15,0)$) |- node [at start] {\numd{1}} node [anchor=east] {$C_{1}$} ($(U.west)+(0,+0.15)$);
	\draw ($(I.north)+(+0.15,0)$) |- node [at start] {\numd{5}} node [anchor=north west] {$E_{1}$} ($(U.west)+(0,-0.15)$);

	\draw ($(U.east)+(0,+0.15)$) -| node [at start] {\numd{2}} node [anchor=west] {$C_{1}$} ($(R.north)+(+0.15,0)$);
	\draw ($(U.east)+(0,-0.15)$) -| node [at start] {\numd{6}} node [anchor=north east] {$R_{1}$} ($(R.north)+(-0.15,0)$);

	\draw (R.south) |- node [at start] {\numd{3}} node [anchor=north] {$C_{2},E_{2}$} (V.east);
	\draw (V.west) -| node [at start] {\numd{4}} node [anchor=north] {$C_{2},R_{2}$} (I.south);

	\node [fit=(U)(V)(I)(R)] {}; 
\end{tikzpicture}
		\caption{MA Delegation Model with Attestation Challenge}
		\label{fig:hema:tp-ma-412}
	\end{subfigure}
	\par\bigskip
	\begin{subfigure}[t]{0.22\linewidth}
		\centering
		\begin{tikzpicture}[
	node distance=1cm,
	-{Latex[scale width=1.5,harpoon]},
	role1/.style={
		draw, rectangle,
		minimum size=1cm, text height=1.5ex, text depth=0.25ex,
		top color=white, bottom color=black!10,
		thick,
	},
	role2/.style={
		draw, rectangle,
		minimum size=1cm,
		minimum height=2cm,
		text height=1.5ex, text depth=0.25ex,
		top color=white, bottom color=black!10,
		thick,
	},
	role3/.style={
		draw, rectangle,
		minimum size=2cm,
		text height=1.5ex, text depth=0.25ex,
		top color=white, bottom color=black!10,
		thick,
	},
	every fit/.style={inner sep=3ex},
]
	\node [role3] (U) at (0,4) {V};
	\node [role3] (I) at (0,0) {I};
	\node [role3] (R) at (4,0) {R};

	\draw ($(I.east)+(0,0.8)$) -- node [at start] {\numd{1}} node [anchor=south] {$C_{1}$} ($(R.west)+(0,0.8)$);
	\draw ($(R.west)+(0,0.5)$) -- node [at start] {\numd{2}} node [anchor=north] {$C_{2},E_{2}$} ($(I.east)+(0,0.5)$);
	\draw ($(I.east)+(0,-0.5)$) -- node [at start] {\numd{7}} node [anchor=south] {$R_{1},R_{2}$} ($(R.west)+(0,-0.5)$);

	\draw ($(I.north)+(-0.8,0)$) -- node [at start] {\numd{3}} node [anchor=east, near start] {$E_{1}$} ($(U.south)+(-0.8,0)$);
	\draw ($(U.south)+(-0.5,0)$) -- node [at start] {\numd{4}} node [anchor=west, near start] {$R_{1}$} ($(I.north)+(-0.5,0)$);

	\draw ($(I.north)+(0.5,0)$) -- node [at start] {\numd{5}} node [anchor=east, near start] {$E_{2}$} ($(U.south)+(0.5,0)$);
	\draw ($(U.south)+(0.8,0)$) --  node [at start] {\numd{6}} node [anchor=west, near start] {$R_{2}$} ($(I.north)+(0.8,0)$);

	\node [fit=(U)(I)(R)] {}; 
\end{tikzpicture}
		\caption{MA Hybrid Model with Attestation Challenge}
		\label{fig:hema:tp-ma-311i}
	\end{subfigure}
	\hfill
	\begin{subfigure}[t]{0.22\linewidth}
		\centering
		\begin{tikzpicture}[
	node distance=1cm,
	-{Latex[scale width=1.5,harpoon]},
	role1/.style={
		draw, rectangle,
		minimum size=1cm, text height=1.5ex, text depth=0.25ex,
		top color=white, bottom color=black!10,
		thick,
	},
	role2/.style={
		draw, rectangle,
		minimum size=1cm,
		minimum height=2cm,
		text height=1.5ex, text depth=0.25ex,
		top color=white, bottom color=black!10,
		thick,
	},
	role3/.style={
		draw, rectangle,
		minimum size=2cm,
		text height=1.5ex, text depth=0.25ex,
		top color=white, bottom color=black!10,
		thick,
	},
	every fit/.style={inner sep=3ex},
]
	\node [role3] (V) at (4,4) {V};
	\node [role3] (I) at (0,0) {I};
	\node [role3] (R) at (4,0) {R};

	\draw ($(I.east)+(0,0.8)$) -- node [at start] {\numd{1}} node [anchor=south] {$C_{1}$} ($(R.west)+(0,0.8)$);
	\draw ($(R.west)+(0,0.5)$) -- node [at start] {\numd{2}} node [anchor=north] {$C_{2}$} ($(I.east)+(0,0.5)$);
	\draw ($(I.east)+(0,-0.5)$) -- node [at start] {\numd{3}} node [anchor=south] {$E_{1}$} ($(R.west)+(0,-0.5)$);
	\draw ($(R.west)+(0,-0.8)$) -- node [at start] {\numd{8}} node [anchor=north] {$R_{1},R_{2}$} ($(I.east)+(0,-0.8)$);

	\draw ($(R.north)+(-0.8,0)$) -- node [at start] {\numd{4}} node [anchor=east, near start] {$E_{1}$} ($(V.south)+(-0.8,0)$);
	\draw ($(V.south)+(-0.5,0)$) -- node [at start] {\numd{5}} node [anchor=west, near start] {$R_{1}$}  ($(R.north)+(-0.5,0)$);

	\draw ($(R.north)+(0.5,0)$) -- node [at start] {\numd{6}} node [anchor=east, near start] {$E_{2}$} ($(V.south)+(0.5,0)$);
	\draw ($(V.south)+(0.8,0)$) -- node [at start] {\numd{7}} node [anchor=west, near start] {$R_{2}$} ($(R.north)+(0.8,0)$);

	\node [fit=(V)(I)(R)] {}; 
\end{tikzpicture}
		\caption{MA Hybrid Model with Attestation Challenge}
		\label{fig:hema:tp-ma-311r}
	\end{subfigure}
	\hfill
	\begin{subfigure}[t]{0.22\linewidth}
		\centering
		\begin{tikzpicture}[
	node distance=1cm,
	-{Latex[scale width=1.5,harpoon]},
	role1/.style={
		draw, rectangle,
		minimum size=1cm, text height=1.5ex, text depth=0.25ex,
		top color=white, bottom color=black!10,
		thick,
	},
	role2/.style={
		draw, rectangle,
		minimum size=1cm,
		minimum height=2cm,
		text height=1.5ex, text depth=0.25ex,
		top color=white, bottom color=black!10,
		thick,
	},
	role3/.style={
		draw, rectangle,
		minimum size=2cm,
		text height=1.5ex, text depth=0.25ex,
		top color=white, bottom color=black!10,
		thick,
	},
	every fit/.style={inner sep=3ex},
]
	\node [role3] (U) at (0,4) {V1};
	\node [role3] (V) at (4,4) {V2};
	\node [role3] (I) at (0,0) {I};
	\node [role3] (R) at (4,0) {R};

	\draw ($(I.east)+(0,0.8)$) -- node [at start] {\numd{1}} node [anchor=south] {$C_{1}$} ($(R.west)+(0,0.8)$);
	\draw ($(R.west)+(0,0.5)$) -- node [at start] {\numd{2}} node [anchor=north] {$C_{2},E_{2}$} ($(I.east)+(0,0.5)$);
	\draw ($(I.east)+(0,-0.5)$) -- node [at start] {\numd{7}} node [anchor=south] {$R_{1},R_{2}$} ($(R.west)+(0,-0.5)$);

	\draw ($(I.north)+(-0.8,0)$) -- node [at start] {\numd{3}} node [anchor=east] {$E_{1}$} ($(U.south)+(-0.8,0)$);
	\draw ($(U.south)+(-0.5,0)$) -- node [at start] {\numd{4}} node [anchor=west] {$R_{1}$} ($(I.north)+(-0.5,0)$);

	\draw ($(I.north)+(0.5,0)$) |- node [at start] {\numd{5}} ($(V.south)+(-0.8,-0.85)$) -- node [anchor=east] {$E_{2}$} ($(V.south)+(-0.8,0)$);
	\draw ($(V.south)+(-0.5,0)$) --  node [at start] {\numd{6}} node [anchor=west] {$R_{2}$} ($(V.south)+(-0.5,-1.15)$) -| ($(I.north)+(0.8,0)$);

	\node [fit=(U)(V)(I)(R)] {}; 
\end{tikzpicture}
		\caption{MA Hybrid Model with Attestation Challenge}
		\label{fig:hema:tp-ma-312i}
	\end{subfigure}
	\hfill
	\begin{subfigure}[t]{0.22\linewidth}
		\centering
		\begin{tikzpicture}[
	node distance=1cm,
	-{Latex[scale width=1.5,harpoon]},
	role1/.style={
		draw, rectangle,
		minimum size=1cm, text height=1.5ex, text depth=0.25ex,
		top color=white, bottom color=black!10,
		thick,
	},
	role2/.style={
		draw, rectangle,
		minimum size=1cm,
		minimum height=2cm,
		text height=1.5ex, text depth=0.25ex,
		top color=white, bottom color=black!10,
		thick,
	},
	role3/.style={
		draw, rectangle,
		minimum size=2cm,
		text height=1.5ex, text depth=0.25ex,
		top color=white, bottom color=black!10,
		thick,
	},
	every fit/.style={inner sep=3ex},
]
	\node [role3] (U) at (0,4) {V1};
	\node [role3] (V) at (4,4) {V2};
	\node [role3] (I) at (0,0) {I};
	\node [role3] (R) at (4,0) {R};

	\draw ($(I.east)+(0,0.8)$) -- node [at start] {\numd{1}} node [anchor=south] {$C_{1}$} ($(R.west)+(0,0.8)$);
	\draw ($(R.west)+(0,0.5)$) -- node [at start] {\numd{2}} node [anchor=north] {$C_{2}$} ($(I.east)+(0,0.5)$);
	\draw ($(I.east)+(0,-0.5)$) -- node [at start] {\numd{3}} node [anchor=south] {$E_{1}$} ($(R.west)+(0,-0.5)$);
	\draw ($(R.west)+(0,-0.8)$) -- node [at start] {\numd{8}} node [anchor=north] {$R_{1},R_{2}$} ($(I.east)+(0,-0.8)$);

	\draw ($(R.north)+(-0.8,0)$) |- node [at start] {\numd{4}} ($(U.south)+(+0.5,-1.15)$) -- node [anchor=east] {$E_{1}$} ($(U.south)+(+0.5,0)$);
	\draw ($(U.south)+(+0.8,0)$) -- node [at start] {\numd{5}} node [anchor=west] {$R_{1}$} ($(U.south)+(+0.8,-0.85)$) -| ($(R.north)+(-0.5,0)$);

	\draw ($(R.north)+(0.5,0)$) -- node [at start] {\numd{6}} node [anchor=east] {$E_{2}$} ($(V.south)+(0.5,0)$);
	\draw ($(V.south)+(0.8,0)$) -- node [at start] {\numd{7}} node [anchor=west] {$R_{2}$} ($(R.north)+(0.8,0)$);

	\node [fit=(U)(V)(I)(R)] {}; 
\end{tikzpicture}
		\caption{MA Hybrid Model with Attestation Challenge}
		\label{fig:hema:tp-ma-312r}
	\end{subfigure}
	\caption{Topological patterns for MA showing the data flow between the Initiator, the Responder, and the Verifier(s). Each of the four rows shows, in order, the Passport Model, the Background-Check Mode, the Delegation Model, and the Hybrid Model. In the first and second columns we use one Verifier and in the third and fourth columns we use two Verifiers. Note the versions with nonces, for freshness, are in the second and fourth columns as well as the entire fourth row. The \hema protocol corresponds to \autoref{fig:hema:tp-ma-111} and \autoref{fig:hema:tp-ma-112} for one Verifier and two Verifiers, respectively.}
	\label{fig:hema:modelos-ma}
\end{figure*}

\autoref{fig:hema:tp-ma-101} is the MA Passport Model where each Attester obtains its own Attestation Report, from the Verifier, and sends it to its peer. \autoref{fig:hema:tp-ma-111} and \autoref{fig:hema:tp-ma-112} depict the versions with nonces with one Verifier and two Verifiers, respectively, and are the reference models discussed throughout \autoref{sec:hema} and in particular in the protocol section (\sref{sec:hema:protocol}). There is no advantage in having parallel messages versus sequential messages, and the number of communication channels required is three with either one Verifier or two Verifiers. This topology has the advantage that is distributes the load in such a way each Attester is responsible for obtaining its own Attestation Report from the Verifier.

\autoref{fig:hema:tp-ma-201} is the MA Background-Check Model where each Attester obtains its peer's Attestation Report from the Verifier. \autoref{fig:hema:tp-ma-211} and \autoref{fig:hema:tp-ma-212} are the versions with nonces with one Verifier and two Verifiers, respectively. In \autoref{fig:hema:tp-ma-211} the Initiator cannot send its Attestation Evidence, $E_{1}$, in the first message to the Responder because creating the Attestation Evidence requires having both nonces and the Initiator receives the Responder's nonce only on the second message.

\autoref{fig:hema:tp-ma-401} is the MA Delegation Model where each Attester sends its Attestation Evidence to the Verifier and the Verifier delivers the Attestation Report to the Attester's peer. The Attesters do not communicate directly with one another during the attestation process. \autoref{fig:hema:tp-ma-411} and \autoref{fig:hema:tp-ma-412} are the versions with nonces with one Verifier and two Verifiers, respectively.

\autoref{fig:hema:tp-ma-311i} and \autoref{fig:hema:tp-ma-311r} are hybrid approaches mixing the Passport Model and the Background-Check Model, and differ in which Attester, the Initiator or the Responder, exchanges all messages with the Verifier. The total number of messages is 7 and 8, respectively, and the amount of channels required is 2 in both cases. However, messages \numd{3}\numd{4} and messages \numd{5}\numd{6} can be exchanged in parallel for the first case and messages \numd{4}\numd{5} and messages \numd{6}\numd{7} can be exchanged in parallel for the second case leading to a reduction of the exchange of messages to 5 and 6, respectively. The process could be made more efficient if both requests to the Verifier, and their respective responses, are batched together, that is, $E_1$ and $E_2$ are sent in a single message and $R_1$ and $R_2$ are received in single message.
\autoref{fig:hema:tp-ma-312i} and \autoref{fig:hema:tp-ma-312r} are the versions with two Verifiers.

\subsubsection{Cost}

\autoref{hema:tab:modelos} compares the cost of the different reference models in regard to the number of messages exchanged when done sequentially and when done in parallel, and the amount of channels required depending on whether there is one Verifier or two Verifiers in the topology.
The number of messages exchanged in parallel are counted per interval where N messages in the same interval count as one message exchange (the number of messages exchanged is still the same, but the time needed for the exchange is reduced due to exchanging messages in parallel).
The number of messages exchanged, whether in sequence or in parallel, is the same with both one Verifier and two Verifiers, in all the reference models.
The agents are connected by bidirectional communication channels that can operate in parallel, but communication within the same channel happens in one direction at a time.

\begin{table}
	\caption{Cost of the topological patterns in \hema.}
	\label{hema:tab:modelos}
	\centering
	\begin{tabulary}{\textwidth}{@{}Lcccc@{}}
		\toprule
		& \multicolumn{2}{c}{\#Messages in} & \multicolumn{2}{c}{\#Channels with} \\
		\cmidrule(lr){2-3} \cmidrule(l){4-5}
		Model & Sequential & Parallel & 1 Verifier &  2 Verifiers \\
		\midrule
		\autoref{fig:hema:tp-ma-101} & 6 & 6 & 3 & 3 \\
		\autoref{fig:hema:tp-ma-201} & 6 & 4 & 3 & 3 \\
		\autoref{fig:hema:tp-ma-401} & 4 & 4 & 2 & 4 \\
		\autoref{fig:hema:tp-ma-111} & 7 & 7 & 3 & 3 \\
		\autoref{fig:hema:tp-ma-211} & 7 & 5 & 3 & 3 \\
		\autoref{fig:hema:tp-ma-411} & 6 & 6 & 2 & 4 \\
		\autoref{fig:hema:tp-ma-311i} & 7 & 5 & 2 & 3 \\
		\autoref{fig:hema:tp-ma-311r} & 8 & 6 & 2 & 3 \\
		\bottomrule
	\end{tabulary}
\end{table}

\section{More on Remote Attestation}
\label{sec:hera:more}

\subsection{Requirements for Attestation Evidence and Attestation Report}
\label{sec:hera:obj}

This section begins with an attestation example, and then lists and describes the minimum set of fields the Attestation Evidence and the Attestation Report artifacts must have. The focus is on the appraisal of both artifacts with regards to the security state of the Attester and not its functionality.

\subsubsection{Attestation Scenarios}

Deciding which claims are mandatory for a cross-TEE attestation scheme requires looking into where the security checks take place, and by whom, since what is essential for one attestation scheme may be superflous for another. As an example, consider the typical system with an Attester, a Verifier, and a Relying Party. The input is a TA image which contains the TA code and initial data, and the public key of the author and their signature over the TA image measurement for authenticity. An entity using a TA usually wants to know whether the TEE is genuine (authentication) and up to date (integrity), and whether the TA is genuine (authentication) and up to date (integrity).
These are separate issues.

The TEE evaluation is best handled by the Verifier who uses endorsements, produced by the Endorser, to ensure the Attestation Evidence produced in the Attester can be trusted, that is, that the RoT of the TEE is valid and it is okay to trust what it is telling the Verifier; and who uses reference values, produced by the Reference Value Provider, to ensure the TEE is up to date.
The TA evaluation can take place at the Attester, at the Verifier, at the Relying Party, or at any combination of these entities; and possibly on other entities in less traditional architectures.
Note that authenticating the TEE does not imply revealing its identity~\cite{Intel:16:EPID} whereas in the case of the TA it is usually best to know who authored the TA image to prevent trust from being placed on a TA produced by a malicious entity.
In our example, the Attester launches a new TA instance, from a TA image, and computes its Image MR. Then, it compares the computed Image MR with the one attached to the TA image to ensure there is a match. Finally, one of the following scenarios can occur, depending on who verifies the author's signature over the Image MR using the author's public key.

\begin{itemize}

	\item In the first scenario, the Attester verifies the author's signature over the Image MR, and sends the claims TA ID, TA security level, TEE ID and TEE security level, to the Verifier along with its own -- the Attester's -- decision on whether the measurement is correct. (Recall the Verifier trusts the TEE because of its RoT.) In such situation, the Verifier does not need to receive the Author MR or the Image MR since this information is verified by the TEE. This could work in an open TEE where anyone can run TAs.
	Such TAs would probably have to run in a sandbox to prevent them from accessing untrusted memory, otherwise a malicious user could run malware within the safety of a TEE while maintaining access to untrusted memory.
	In this scenario, the Verifier can check whether the TEE is up to date using the TEE ID and the TEE security level, and whether the TA is up to date using the TA ID and the TA security level, while trusting the author's signature has already been verified by the Attester.

	\item In the second scenario, the Attester sends the computed Image MR, along with the author's signature over Image MR, and the Author MR to the Verifier who checks whether the signature is correct. The Verifier has the public keys of the authors and can verify the signatures it receives.

	\item In the third scenario, the Verifier checks whether the TEE is genuine and up to date, but forwards the Image MR, the Author MR, and the author's signature over the Image MR to the Relying Party who checks whether the signature is correct. The Relying Party has the public keys of the authors it considers relevant for its use case enabling it to run its own checks.

\end{itemize}

Therefore different parts of the attestation process can be conducted by different roles, or even, redundantly, repeated by different roles, taking into account what information is available to each of those roles. Certain roles are more likely to have all the information needed to make decisions regarding certain items. The Verifier is better placed to evaluate the TEE status since it has information regarding the RoT of the TEE, and the Relying Party is better placed to evaluate the TA status since it is the entity using the attestation results to reach a trust decision.

\subsubsection{Required Fields}

\autoref{tab:hema:ra} lists the minimum set of fields that must be present in the Attestation Evidence and the Attestation Report artifacts to have a working, and interoperable, attestation scheme that is not limited to one type of TEE and one Verifier. A TEE can add additional claims (see \sref{sec:hera:obj:more}) at the end of this section) about the environment and add additional fields. This flexibility is necessary due to the heterogeneity of TEEs available.

\begin{table}
	\centering
	\caption{Minimum requirements for the Attestation Evidence and the Attestation Report.}
	\label{tab:hema:ra}
	\begin{tabulary}{\linewidth}{@{}llL@{}}
		\toprule
		& Field & Description \\
		\midrule
		\multicolumn{2}{@{}l@{}}{\textbf{Attestation Evidence}} & \\
		& Version & The version of the Attestation Evidence artifact. \\
		& User data & An optional bitstring provided by the user. \\
		& Claims & Set of claims made by the Attester. \\
		& Proof & Attester's signature over Attestation Evidence. \\
		\multicolumn{2}{@{}l@{}}{\textbf{Claims}} & \\
		& TEE ID & Identifies this type of TEE. \\
		& TA ID & Identifies this specific TA. \\
		& TEE security level & The security version of the TEE. \\
		& TA security level & The security version of the TA. \\
		\midrule
		\multicolumn{2}{@{}l@{}}{\textbf{Attestation Report}} & \\
		& Version & The version of the Attestation Report artifact. \\
		& Evidence & The Attestation Evidence produced by the Attester. \\
		& Claims & Set of claims made by the Verifier. \\
		& Signature & The Verifier's signature over the Attestation Report. \\
		\multicolumn{2}{@{}l@{}}{\textbf{Claims}} & \\
		& Verifier ID & Identifies the Verifier producing the Attestation Report. \\
		& Trustworthiness & Whether the TA and TEE are considered legitimate by the Verifier. \\
		\bottomrule
	\end{tabulary}
\end{table}

The \emph{version} of the Attestation Evidence, and of the Attestation Report, is used by both the producer and the consumer of the respective artifact to know which fields are mandatory and which fields are optional, and how the data is structured. This enables the artifacts to evolve while remaining backward compatible since it is unlikely all systems are updated simultaneously.

\subsubsection{Attestation Evidence}

The \emph{user data} is an optional bitstring provided by the caller of the function. This field is typically used to bind nonces and keying material to the Attestation Evidence and to the Attestation Report. The data itself is commonly hashed and the resulting digest stored in the user data field and thus bound to the Attestation Evidence.

The \emph{proof} in the Attestation Evidence is the Attester's signature over the Attestation Evidence, minus the proof field itself. The Verifier has the corresponding public key and is able to verify the signature. In some cases, the Attester and the Verifier may, instead of a key pair, share symmetric keys with the proof field holding the resulting authenticity computation, for example, a MAC over the Attestation Evidence. The private key or the secret key is stored securely in the Attester and accessible only to the TEE. The possession of this key gives the Verifier confidence the Attester's TEE is real and not being emulated or simulated.

There are four mandatory claims for the Attestation Evidence: the IDs and security levels of both the TEE and the TA. The \emph{TEE ID} identifies the TEE in use, for example, SGX. This is of particular importance when a Verifier issues Attestation Reports for TEEs of different types. The \emph{TA ID} identifies the TA in use. A software house, for example, could have multiple TAs each of which would have its own TA ID. Neither TEE ID nor TA ID change when there are functionality or security updates. These values are fixed. The \emph{TEE security level} and the \emph{TA security level} contain the current version of the security state of the TEE and the TA, respectively. The Verifier uses these four fields to find whether the TEE and the TA are up to date.

These four claims are the minimum to ensure the TEE and the TA are genuine and up to date.

\subsubsection{Attestation Report}

The Attestation Report contains the Attestation Evidence, sent by the Attester, in the \emph{evidence} field. This allows the consumer of the Attestation Report to use the Attestation Evidence, in particular the claims, during its appraisal.

The \emph{signature} in the Attestation Report is the Verifier's signature over the Attestation Report, minus the signature field itself. The Relying Party has the corresponding public key, or a signature chain rooted on a private key controlled by the Verifier, enabling its verification. This gives the Relying Party confidence in the origin of the Attestation Report and that the information contained within is genuine.

There are two mandatory claims for the Attestation Report: The Verifier ID and the decision made regarding the trustworthiness of the Attester. The \emph{Verifier ID} uniquely identifies the Verifier. This could be, for example, its public key. The Verifier ID is needed to support multiple Verifiers in our attestation scheme. The \emph{trustworthiness} field contains the decision made by the Verifier, after its appraisal of the Attestation Evidence, regarding whether the Attester is legitimate. The Relying Party uses this field during its appraisal of the Attestation Report.

\subsubsection{Additional Claims}
\label{sec:hera:obj:more}

Two claims that are often used in attestation but which are not mandatory in our model are the Author MR and the Image MR. The \emph{Author MR} identifies the entity responsible for the TA and the \emph{Image MR} is a cryptographic hash over the TA code and initial data. These claims are optional in the Attestation Evidence and the Attestation Report because they are not essential to the appraisal performed by the Verifier and the Relying Party.

Other claims include the TEE and the TA versions and configurations. The \emph{TEE version} and the \emph{TA version} show the functionality level of the TEE and of the TA, respectively. The TEE and the TA may receive functionality upgrades and patches over time but this does not necessarily imply a security level change. An example would be a performance optimization. The \emph{TEE configuration} and the \emph{TA configuration} show how the TEE and the TA are configured. The TEE and the TA can be launched with different amounts of resources and permissions, for example, and the Relying Party may want to use this information when appraising the Attestation Evidence.

\subsection{Functions}
\label{sec:hera:fun}

\autoref{tab:hera:fun} shows the functions used by the \hera protocol which is presented in the next section (\sref{sec:hera:proto}). In addition, the \hera protocol also uses some of the functions defined in \autoref{tab:hema:fun} for the \hema protocol.

\begin{table*}
	\centering
	\caption[Functions used by the \hera protocol]{Functions used by the \hera protocol.}
	\label{tab:hera:fun}
	\begin{tabulary}{\textwidth}{@{}llL@{}}
		\toprule
		Caller & Function & Description \\
		\midrule

		Attester & $\rfun{attest}{m}{\var{AE}}$ & Creates Attestation Evidence, $\var{AE}$, based on a set of claims and an optional custom message $m$. \\

		Verifier & $\rfun{verify}{\var{AE}, m, \var{VeP}}{\var{AR}}$ & Verifies the claims in the Attestation Evidence, $\var{AE}$, and produces an Attestation Report, $\var{AR}$, as a result. The bitstring $m$ is optional. \\

		\bottomrule
	\end{tabulary}
\end{table*}

The function \fun|ra| is called by the Attester to trigger the RA protocol, depicted in \autoref{fig:hera:sd:1}. The bitstring $m$, received as input, is optional and is contained within the Attestation Evidence returned as output of \fun|ra|. The functions \fun|attest| and \fun|verify| are called internally by the Attester and the Verifier, respectively, as part of the \fun|ra| invocation.

The function \fun|validate| is called by the Relying Party to find whether the Attester is trustworthy and can be trusted. The Attestation Report contains information about the Attester and a decision, made by the Verifier, on whether the Attester is trustworthy. The Validation Policy contains the rules dictating what a trusted Attester looks like, and, in conjunction with the Attestation Report and the optional bitstring, helps the Relying Party make a decision on whether the Attester can be trusted. The Relying Party may find an Attester trustworthy but still decide not to trust it. For example, a bank application could be considered trustworthy by the Verifier and still not be trusted by the Relying Party because the TA belongs to a different bank of the one expected by the Relying Party.

\subsection{\hera Protocol}
\label{sec:hera:proto}

This section describes the \hera (Heterogeneous Remote Attestation) protocol using a top-down approach. The \hera protocol enables cross-platform RA. First we describe the \emph{use} of the \lstinline|ra| primitive, from a high-level perspective, and then \emph{how} the primitive looks internally.

\begin{figure}
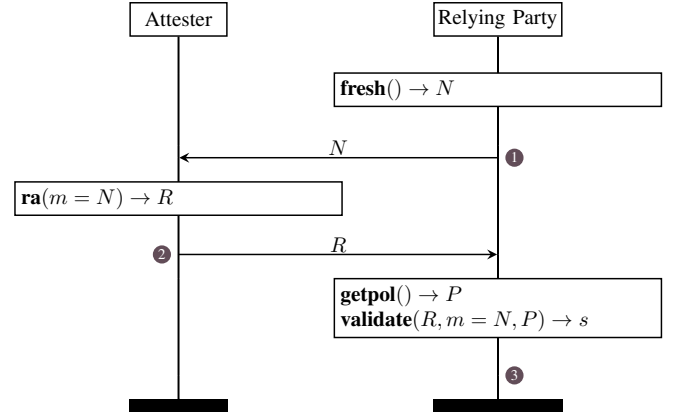

	\centering
	\includestandalone[scale=0.8]{sd-hera-1}
	\caption[Invocation of the RA primitive by the Attester]{High-level view of the \hera protocol showing the invocation of the \fun|ra| primitive by the Attester. Nonce $N$ tells the Relying Party that the Attestation Report $R$ is fresh.}
	\label{fig:hera:sd:1}
\end{figure}

\autoref{fig:hera:sd:1} shows a sequence diagram, between an Attester and a Relying Party, exemplifying the invocation of the \lstinline|ra| function. The protocol steps are the following:

\begin{enumerate}
	\item \label{fig:hera:sd:1:step-1} The Relying Party generates and sends a fresh nonce, $N$, to the Attester.

	\item The Attester invokes the \lstinline|ra| function using $N$ as input and replies to the Relying Party with the resulting Attestation Report, $R$.

	\item The Relying Party obtains a Validation Policy, $P$, provisioned by the Relying Party Owner, and appraises $R$ using $P$ as a reference for trustworthiness and the previously sent nonce $N$ as a reference for freshness.
\end{enumerate}

\autoref{fig:hera:sd:0} shows the sequence diagram for the \hera protocol between an Attester and a Verifier, illustrating how the \lstinline|ra| primitive works internally. The protocol steps are the following:

\begin{enumerate}
	\item The Attester invokes the \lstinline|attest| function to collect and sign a set of claims about itself resulting in Attestation Evidence, $E$. A custom message, $m$, is given to \lstinline|attest| as input to be included in the Attestation Evidence. The Attester sends $E$ to the Verifier.

	\item The Verifier validates the Attestation Evidence, $E$, received from the Attester. In particular, it (i) verifies the signature of the Attesting Environment over $E$, and (ii) checks whether $E$ corresponds to a legitimate instance running on an up-to-date TEE. Then, the Verifier produces an Attestation Report, $R$, and sends $R$ to the Attester.

	\item The Attester now has an Attestation Report, $R$, certified by the Verifier, that it can use to attest itself before another party. $R$ includes the contents of $E$, namely the custom message $m$ which the Attester used as input to the \lstinline|attest| function.
\end{enumerate}

The custom message, $m$, is traditionally used to exchange public keys or nonces. The Attestation Evidence, $E$, is signed by the Attester thus binding the keying material to the TA instance. In the previous sequence diagram, in \autoref{fig:hera:sd:1}, $m$ contains a nonce generated by the Relying Party ensuring both the Attestation Evidence and the Attestation Report are fresh.

\begin{figure}
	\centering
	\includestandalone[scale=0.8]{sd-hera-0}
	\caption[The \hera protocol]{The \hera protocol.}
	\label{fig:hera:sd:0}
\end{figure}

Upon a successful conclusion of the attestation protocol, the Relying Party can continue by providing service to the Attester.

\section{Output of Bugged \hema Model}
\label{app:hema:miiisf}

\autoref{lst:hema:formal:output:miiisf} shows the output of running a version of the \hema model and proofs through Tamarin. Notice how the sanity checks fail, except for \lstinline|final_7| which is slightly different from the other as explained in \autoref{sec:hema:formal:model:sanity}, but the proofs all pass successfully. Without any sanity checks it is possible for an error, which in this case was introduced purposefully, to not be caught.

\begin{lstlisting}[language=tamarin-output, style=tese, caption={Output of running the proofs on a model of the \hema scheme with a bug introduced to illustrate how a missing variable can lead to proofs misleadingly holding.}, label=lst:hema:formal:output:miiisf, morekeywords={hash}]
  final_1 (exists-trace): falsified - no trace found (5 steps)
  final_2 (exists-trace): falsified - no trace found (5 steps)
  final_3 (exists-trace): falsified - no trace found (5 steps)
  final_4 (exists-trace): falsified - no trace found (5 steps)
  final_5 (exists-trace): falsified - no trace found (5 steps)
  final_6 (exists-trace): falsified - no trace found (5 steps)
  final_7 (all-traces): verified (5 steps)
  final_8 (exists-trace): falsified - no trace found (5 steps)
  secrecy (all-traces): verified (10 steps)
  aliveness_Initiator (all-traces): verified (2 steps)
  aliveness_Responder (all-traces): verified (3 steps)
  weak_Initiator (all-traces): verified (2 steps)
  weak_Responder (all-traces): verified (3 steps)
  nia_Initiator (all-traces): verified (2 steps)
  nia_Responder (all-traces): verified (3 steps)
  ia_Initiator (all-traces): verified (2 steps)
  ia_Responder (all-traces): verified (3 steps)
\end{lstlisting}

\end{document}


\begin{tikzpicture}[estado estilo]

	\pic (S0) {estado={S0}{`start'}{
		Create TA instance for Initiator\\
	}};
	\pic (I1) [below=of S0-g.south -| S0-a] {estado={I1}{}{
		Generate ANonce\\
		Generate key pair\\
		Send M1 = ANonce, APub\\
	}};
	\pic (I3) [below=of I1-g.south -| I1-a] {estado={I3}{}{
		Let AHash = hash(ANonce+BNonce+APub+BPub)\\
		Validate BReport\\
		AReport = ra(AHash)\\
		Send M3 = AReport\\
	}};
	\pic (F0) [below=of I3-g.south -| I3-a] {estado={F0}{}{
		Compare session keys\\
	}};

	\draw [estado seta] (S0-g) -- node[right] {} (I1-g);
	\draw [estado seta] (I1-g) -- node[right] {Receive M2} (I3-g);
	\draw [estado seta] (I3-g) -- node[right] {} (F0-g);

	\node (A0) [above=of S0-g, anchor=center] {\huge\faServer};
	\draw [estado seta] (A0) -- (S0-g);

\end{tikzpicture}


\begin{tikzpicture}[estado estilo]

	\pic (S0) {estado={S0}{`start'}{
		Create TA instance for Responder\\
	}};
	\pic (R2) [below=of S0-g.south -| S0-a] {estado={R2}{}{
		Generate BNone\\
		Generate key pair\\
		Let BHash= hash(ANonce+BNonce+APub+BPub)\\
		BReport=ra(BHash)\\
		Send M2 = BNonce, BPub, BReport\\
	}};
	\pic (R4) [below=of R2-g.south -| R2-a] {estado={R4}{}{
		Validate AReport\\
	}};
	\pic (F0) [below=of R4-g.south -| R4-a] {estado={F0}{`end'}{
		Compare session keys\\
	}};

	\node (B0) [above=of S0-g, anchor=center] {\huge\faServer};
	\draw [estado seta] (B0) -- node[right] {} (S0-g);

	\draw [estado seta] (S0-g) -- node[right] {Receive M1} (R2-g);
	\draw [estado seta] (R2-g) -- node[right] {Receive M3} (R4-g);
	\draw [estado seta] (R4-g) -- node[right] {} (F0-g);

\end{tikzpicture}